\documentclass[%
 reprint,
superscriptaddress,
 amsmath,amssymb,
 aps,
]{revtex4-2}

\usepackage{graphicx}% Include figure files
\usepackage{dcolumn}% Align table columns on decimal point
\usepackage{bm}% bold math
\usepackage{xcolor}
\begin{document}

\preprint{APS/123-QED}

\title{Long-living dark coherence brought to light by magnetic-field controlled photon echo}

\author{I.~A.~Solovev}
\author{I.~I.~Yanibekov}
\author{Yu.~P.~Efimov}
\author{S.~A.~Eliseev}
\author{V.~A.~Lovcjus}
\affiliation{
 Saint Petersburg State University, 1 Ulyanovskaya Str., Saint Petersburg, 198504, Russia
}
\author{I.~A.~Yugova}
\author{S.~V.~Poltavtsev}
\affiliation{
 Spin Optics Laboratory, Saint Petersburg State University, 1 Ulyanovskaya Str., Saint Petersburg, 198504, Russia
}
\author{Yu.~V.~Kapitonov}
 \email{yury.kapitonov@spbu.ru}
\affiliation{
 Saint Petersburg State University, 1 Ulyanovskaya Str., Saint Petersburg, 198504, Russia
}

\date{\today}

\begin{abstract}
Larmor precession of the quasiparticle spin about a transverse magnetic field leads to the oscillations in the spontaneous photon echo signal due to the shuffling of the optical coherence between optically accessible (bright) and inaccessible (dark) states. Here we report on a new non-oscillating photon echo regime observed in the presence of non-equal dephasing rates of bright and dark states. This regime enables the observation of the long-living dark optical coherence. As a simple mechanical analogy, we suggest a charged particle moving in the magnetic field through the  medium with anisotropic viscous friction. We demonstrate the dark coherence retrieval in the spontaneous photon echo from excitons in the InGaAs/GaAs quantum well.
\end{abstract}

\maketitle

The coherence manipulation is the cornerstone of the optical quantum information processing~\cite{Kim, Biolatti, Li, Kroutvar}. The optical coherence could be addressed and stored in ensembles of resonant systems by the photon echo (PE) protocol, in the simplest cases realized by two (spontaneous PE) or three (stimulated PE) successive laser pulses exciting the ensemble~\cite{Bits, Lvovsky, Hammerer, Polt2018PE}. An important step towards applications utilizing such coherence manipulation in the ultrafast way was the observation of PE from ensembles of excitons and their complexes in epitaxial heterostructures based on GaAs~\cite{Sol2018InGaAs, Sal2017InGaAsQD, Polt2016InGaAsQD}, CdTe~\cite{Polt2017Rabi, Sal2017PRX, Lang2014NatPh, Lang2012PRL}, and wide-band gap ZnO~\cite{Sol2018ZnO, Polt2017ZnO} and GaN~\cite{Polt2018GaN} semiconductors. PE is also actively studied in novel excitonic materials: transition metal dichalcogenide (TMDC) monolayers~\cite{TMDC2016, KaspTMDC2016}, and  halide perovskites~\cite{March2016, Bohn2018, March2017, March2019}.

The additional degree of coherence control is provided by the magnetic-field manipulation of spins of charge carriers constituting the excitons and their complexes. The coupling between the spin states is achieved by the transverse magnetic field involving electron and hole spins in Larmor precession. It allows the periodical optical coherence shuffling between optically accessible and inaccessible states leading to the oscillations in spontaneous PE signal. Such magnetic-field PE control was demonstrated in the trion system in a CdTe/CdMgTe quantum well (QW) \cite{Polt2020PRB, Lang2014NatPh, Lang2012PRL, Sal2017PRX}. It is important to note that in these works the two states involved in the shuffling are equivalent trion states with the same phase relaxation (dephasing) rates, and the accessibility of states is determined by the polarization of the corresponding optical transitions.

Alongside this, there are optically inaccessible states, transitions to which are prohibited by the selection rules. Examples of such states are excitons with the total angular momentum projection $S = \pm 2$ in GaAs- and CdTe-based QWs~\cite{Polt2019Sci}. Prohibition of the interaction with light gave these states the name {\it dark excitons} as opposed to the {\it bright} ones with $S = \pm1$. Application of the external magnetic field makes it possible to observe footprints of dark excitons in non-coherent optical signals, as was done for QWs~\cite{Glasberg1999}, quantum dots~\cite{Bayer2000} and carbon nanotubes~\cite{Zaric2004}. Analogous optically inaccessible states were studied also in TMDC monolayers~\cite{Lu2020, Zhang2017, Tang2019, Zhou2017}. Still the coherent optical properties of dark excitons were little studied~\cite{Ikeuchi2003,Tomoda2010}, and polarization shuffling between states was not addressed at all.

The main distinguishing point in the context of the polarization shuffling between bright and dark excitonic states is the difference of dephasing rates of {\it bright} and {\it dark} coherences. In this paper we will show that this difference leads to quite unexpected results in spontaneous PE with magnetic field, and opens the way to retrieve the long-living dark coherence in optical signal. We will begin with the theoretical consideration of this problem, and then we will demonstrate the predicted dark optical coherence retrieval in the spontaneous PE experiment with excitons in the InGaAs/GaAs QW.

\begin{figure*}
\includegraphics[width=0.85\linewidth]{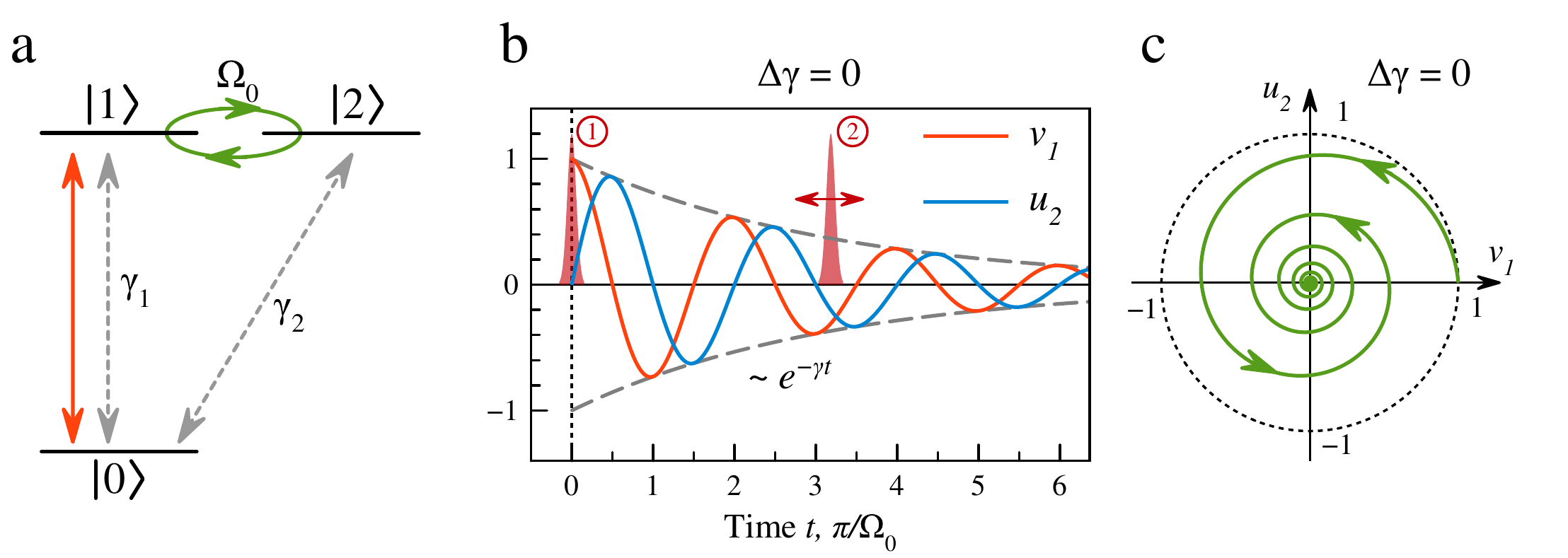}
\caption{\label{Fig1} (a) Energy diagram of the three-level system with the optically addressed transition $|0\rangle \leftrightarrow |1\rangle$ (red arrow) and different dephasing rates for $|0\rangle \leftrightarrow |1\rangle$ and $|0\rangle \leftrightarrow |2\rangle$ polarizations (gray arrows). States $|1\rangle$ and $|2\rangle$ are coupled with the precession frequency $\Omega_0$. (b) Evolution of elements $v_1$ (red curve) and $u_2$ (blue curve) after the action of the $\frac{\pi}{2}$-pulse (1) for the case $\Delta \gamma = 0$. The pulse (2) is scanned. Dashed curve shows the decay of the elements envelope $\sqrt{v^2_1+u^2_2}$. (c) Same evolution shown as the phase portrait on $(v_1,u_2)$-plane for the case $\Delta \gamma = 0$.}
\end{figure*}

The minimal system needed to demonstrate the coherence shuffling consists of three levels schematically shown in Fig.~\ref{Fig1}(a). The ground state $|0\rangle$ and the bright excited state $|1\rangle$ are coupled by the light. The dark excited state $|2\rangle$ could not be accessed by the same optical excitation due to the selection rules. At the same time states $|1\rangle$ and $|2\rangle$ are coupled with the precession frequency $\Omega_0$ (i.e. by the magnetic field). The coherencies $|0\rangle \leftrightarrow |1\rangle$ and $|0\rangle \leftrightarrow |2\rangle$ have different phenomenological dephasing rates $\gamma_1$ and $\gamma_2$ respectively. We will assume the case of the long-living dark coherence ($\gamma_2 \le \gamma_1$). Below we will describe the main stages of the model development only. The strict Lindblad equation solution is provided in~\cite{SMA}.

In the spontaneous PE experiment an inhomogeneously broadened ensemble of three-level systems is excited by two laser pulses separated by the time delay $\tau$. The evolution of one system is described by the density matrix~$\rho$. After the action of the $\frac{\pi}{2}$-pulse the system is driven into the coherent superposition of $|0\rangle$ and $|1\rangle$ states represented by the non-diagonal element of the density matrix $\rho_{01}$. After the pulse action the system starts to evolve in time with the coherencies oscillating between $|0\rangle \leftrightarrow |1\rangle$ and $|0\rangle \leftrightarrow |2\rangle$.
The second $\pi$-pulse rephases the ensemble. The PE signal zero is observed when there is no polarization in the optically accessible state ($\rho_{01} = 0$) at the arrival time of the rephasing $\pi$-pulse. We are interested in the behavior of the PE decay signal, that is, the dependence of the spontaneous PE amplitude observed at the time $t=2 \tau$ on the $\tau$. We will denote this signal as $P_{PE}$. Obviously in the absence of the precession ($\Omega_0 = 0$) the PE signal has exponential decay with $\tau$ with the rate $2 \gamma_1$ ($P_{PE} \sim e^{-2 \gamma_1 \tau}$).

Let's try to build an intuitive picture for the PE decay in the $\Omega_0 \ne 0$ case. At the first glance $P_{PE}$ is expected to have temporal oscillations with frequency $\Omega_0$ due to the polarization shuffling between states~\cite{Lang2012PRL}. The next intuitive step will be to consider the fact that the polarization persists the same time in both excited states, which leads to the supposed average PE decay rate $2 \left( \frac{\gamma_1+\gamma_2}{2} \right)$. We will show that, with minor corrections, this guess is true only for the precession frequencies being higher than the dephasing rates difference. However at lower $\Omega_0$ this intuitive approach is no more valid, and full consideration of the problem is needed.

In order to demonstrate the counter-intuitive discrepancies we will start with the most characteristic property of the expected oscillatory PE decay behaviour -- the appearance of periodical PE signal zeroes. 

The theoretical consideration of this problem shows that the temporal evolution of the coherence $|0\rangle \leftrightarrow |1\rangle$ (and related optical polarization)  is coupled only with the coherence $|0\rangle \leftrightarrow |2\rangle$, and the following simple differential equation could be written for the temporal evolution of the system \cite{SMA}:

\begin{equation}
    \label{eq_xy}
    \left\{
        \begin{array}{l}
        \dot{v}_1 = \phantom{-} \Omega_0 u_2 - \gamma_1 v_1 \\  
        \dot{u}_2 = - \Omega_0 v_1 - \gamma_2 u_2
        \end{array}
    \right.,
\end{equation}

where optical Bloch equation elements are introduced in the following way: $u_2 = 2 {\rm Re} (\rho_{02}$), $v_1 = -2 {\rm Im} (\rho_{01})$. Equation~$\ref{eq_xy}$ has a simple mechanical analogy of a charged particle precessing in the magnetic field in a medium with anisotropic viscous friction~\cite{SMB}.

The $\frac{\pi}{2}$-pulse action leads to $v_1(0) = 1$, $u_2(0)=0$. Since $P_{PE} \sim | v_1(\tau) |^2$, tracking $v_1(\tau_k) = 0$ will lead to the set of PE signal zeroes $\tau_k$.

First we will consider the already known simple case with equal dephasing rates $\gamma_1 = \gamma_2$ $\Delta \gamma = 0$. Equation $\ref{eq_xy}$ gives the following solution: $v_1(t)=e^{-\gamma t} \cos (\Omega_0 t); u_2(t) = e^{-\gamma t} \sin (\Omega_0 t)$, where the  $\gamma = \frac{\gamma_1 + \gamma_2}{2}$ is the average dephasing rate. Corresponding oscillations of elements $v_1$ and $u_2$ with shifted phases and exponential decay of the envelope are depicted in Fig.~\ref{Fig1}(b). PE signal has infinite set of zeroes: $\tau_k = \frac{\frac{\pi}{2} + \pi  k}{\Omega_0}$, $k \in Z$. This oscillatory behaviour could be represented as a phase portrait in the ($v_1, u_2$)-plane giving the shrinking spiral with ($O,u_2$)-axis intersections corresponding to PE zeroes (Fig.~\ref{Fig1}(c)). For the PE decay one can get the following simple expression: $P_{PE} \sim e^{-2 \gamma \tau} \cos^2 \Omega_0 \tau$. Such oscillations between two optical states was observed in the co-circular excitation and detection geometry for the spontaneous PE from trions in CdTe/CdMgTe quantum well subject to the transverse magnetic field~\cite{Lang2012PRL, Polt2020PRB} where both involved states had equal dephasing rates.

\begin{figure}
\includegraphics[width=\linewidth]{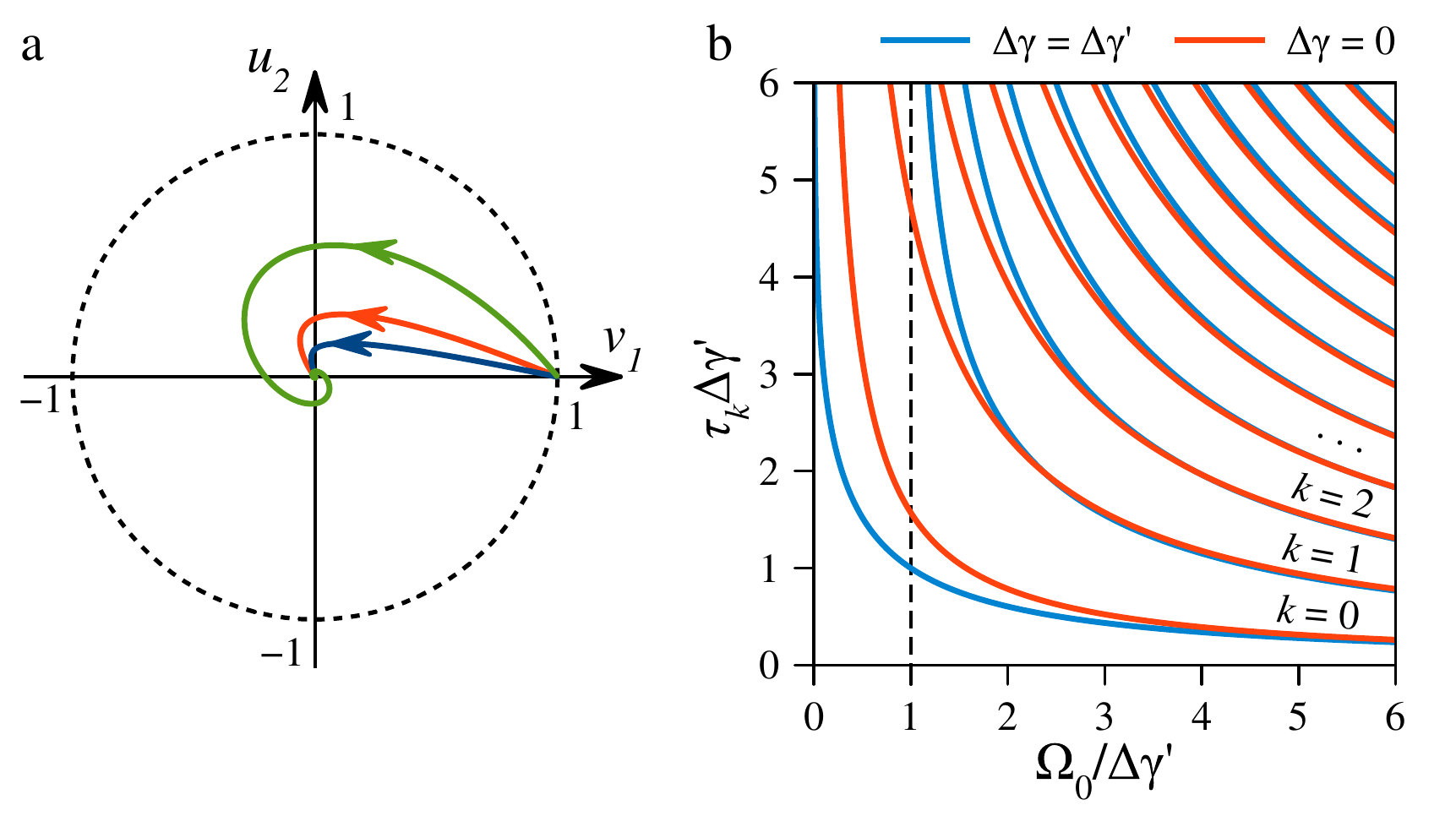}
\caption{\label{Fig_Spirals}
(a) Phase portraits for oscillatory ($\Omega_0 = 3 \Delta \gamma$, green curve), critical ($\Omega_0 = \Delta \gamma$, red curve) and non-oscillatory ($\Omega_0 = 0.5 \Delta \gamma$, blue curve) regimes. $\Delta \gamma / \gamma = 0.7$. 
(b) PE signal zeroes $\tau_k$ for $\Delta \gamma = \Delta \gamma'$ (blue curves) and $\Delta \gamma = 0$ (red curves) plotted in dimensionless coordinates.
}
\end{figure}

The non-zero difference in dephasing rates $\Delta \gamma = \frac{\gamma_1 - \gamma_2}{2}$  leads to a significant modification of the solution of Eq.$\ref{eq_xy}$. This solution branches into three different regimes shown in phase portraits in Fig.~\ref{Fig_Spirals}(a). For the oscillatory regime at $\Omega_0 > \Delta \gamma$ the oscillatory behavior similar to the described above is observed with little corrections of the frequency and phase (\cite{SMA}, Eq.~(S11)). However at the critical precession frequency $\Omega_0 = \Delta \gamma$ a new non-oscillatory regime emerges. In this regime the PE signal goes through zero only once and the phase portrait changes from spiral to infinite ''falling'' to the $(0,0)$ point (Fig.~\ref{Fig_Spirals}(a), red and blue curves). Thus, an infinite set of zeros $\tau_k$ in the oscillatory regime is replaced by a single $\tau_0$ zero in critical and non-oscillatory regimes, which completely breaks the intuitive picture discussed above. Fig.~\ref{Fig_Spirals}(b) shows the PE signal zeroes for the $\Delta \gamma = 0$ and $\Delta \gamma \ne 0$ cases (\cite{SMA}, Eq.~(S9)). Accounting of the dephasing rates difference modifies the expression for the PE decay leading to the general solution 
$P_{PE}
    \sim
    e^{-2 \gamma \tau}
    \left|
        \cos \Omega \tau
        -
        \frac{\Delta \gamma}{\Omega}
        \sin \Omega \tau
    \right|^2
$, where the effective precession frequency $\Omega = \sqrt{\Omega_0^2 - \Delta \gamma^2}$ is introduced. 

It should be noted that for the oscillatory regime ($\Omega_0 > \Delta \gamma$), the above intuitive considerations turn out to be close to the truth: the envelope of oscillations decays with an average dephasing rate $\gamma$ with the doubling of the decay time in the limiting $\gamma_2 = 0$ case. Quite unexpected result could be seen in the non-oscillatory regime ($\Omega_0 < \Delta \gamma$): after the first and only signal zero the PE begins to decay slower, and at $\Omega_0 \rightarrow 0$ the decay rate at long delays tends to the dark coherence dephasing rate $\gamma_2$ (\cite{SMA}, Eq.~(S12)). At moderate $\Omega_0$ the dramatic increase of the decay time could be observed. In such a way the long-living dark coherence could be revealed.

In order to demonstrate these regimes and the ability to address the long-living dark coherence we have carried out the experimental study of the spontaneous PE from excitons in InGaAs/GaAs QW.

\begin{figure*}
\includegraphics[width=\linewidth]{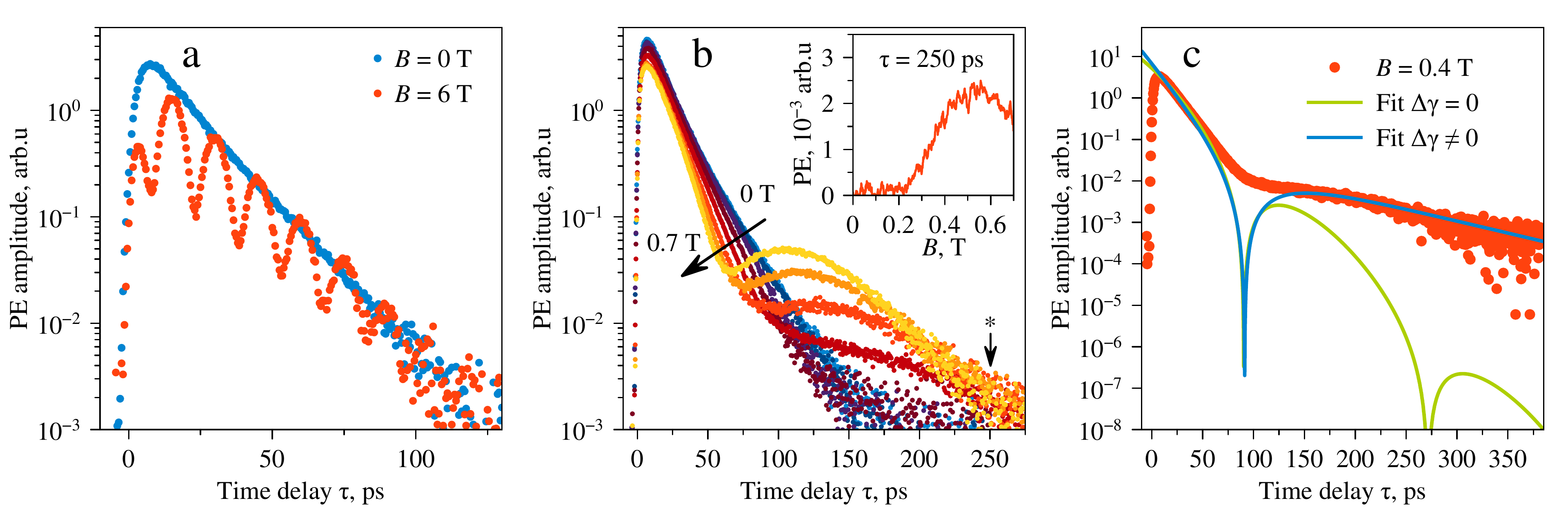}
\caption{\label{Fig_Data} (a) PE decays without magnetic field (blue dots) and at 6~T (red dots). (b) PE decays at magnetic fields from 0~T to 0.7~T with 0.1~T step. Inset shows the PE signal amplitude dependence at $\tau = 250$~ps (marked with asterisk) on magnetic field $B$. (c) Experimental PE decay for $B = 0.4$~T (red dots), fit with equal dephasing rates ($\Delta \gamma = 0$, green solid curve) and different dephasing rates ($\Delta \gamma \ne 0$, blue solid curve).}
\end{figure*} 

The sample P551 was grown on the (100) oriented GaAs substrate by the molecular beam epitaxy and consist of 1~$\mu$m GaAs buffer layer and a 3~nm In$_{0.03}$Ga$_{0.97}$As QW followed by the 170~nm GaAs capping layer.

The PE experiment was carried out with the time-resolved four-wave mixing technique. The sample was kept at 1.5~K. The transverse magnetic field up to 6~T could be applied to the sample. The sample was excited by two successive laser pulses linearly polarized along the magnetic field direction. PE signal was collected in the reflection geometry and detected by the cross-correlation with one more pulse from the same laser source. Cross-correlation measurements opens a way to detect the PE electromagnetic field amplitude with temporal resolution. Ti:Sapphire Spectra Physics Tsunami laser was used as a source of $2$~ps laser pulses with the repetition rate $81$~MHz. Full width at half maximum of the laser pulse spectra was around $1$~meV, which makes possible to resonantly excite the excitonic transition without involving other exciton complexes. More details on the experimental technique could be found in~\cite{Polt2020PRB, Sol2018InGaAs}.

Experiments were carried out on the QW heavy-hole exciton transition. Reflectivity measurements~\cite{Polt2014Brew, Polt2010Brew} were used to determine its parameters at $T=10$~K: resonance position $1.510$~eV, radiative width $34$~$\mu$eV, and the non-radiative broadening $71$~$\mu$eV.

In order to describe the energy diagram of the heavy-hole excitonic transition in QW a five-level system could be introduced consisting of the ground state, two bright excitons with $S = \pm 1$ and two dark excitons with $S = \pm 2$. The dark excitons do not interact with light, and bright excitons with $S=+1$ ($-1$) could be addressed by the $\sigma^+$ ($\sigma^-$) polarized light. In transverse magnetic field the Larmor precession of electron and heavy hole spins takes place. These precessions couple the bright and dark exciton states opening the way to transfer the optical coherence between them. Still the complicated spin dynamics could lead to complex polarization-dependent behaviour.

A significant simplification that allows us to reduce this system to three levels is the representation of exciton system in the linear polarizations basis \cite{SMC}. We will denote by H and V the polarizations along and perpendicular to the transverse magnetic field $\vec{B}$ correspondingly. 
In the linear polarizations basis and the case of the H-polarized excitation pulses the evolution of the H-polarized excitonic states is decoupled from V-polarized states. Dark and bright H-polarized excitonic states are coupled by the magnetic field with the precession frequency $\Omega_H = \frac{\Omega_e + \Omega_h}{2}$, where $\Omega_e$ and $\Omega_h$ are the Larmor precession frequencies for electron and hole correspondingly. Thus in this polarization geometry we could apply the above developed three-level theoretical model to the system consisting of the ground state, H-polarized bright and dark exciton states and H-polarized excitation and detection.

Fig.~\ref{Fig_Data}(a) shows the PE decay measured without the magnetic field (blue dots). It shows the exponential decay from which the bright exciton dephasing time $T_{2b} = 1/\gamma_1 = 29$~ps could be extracted. At magnetic field $B = 6$~T (red dots) PE signal oscillations could be observed with frequency $0.1$~ps$^{-1}$. This signal clearly demonstrates the oscillatory regime ($\Omega_0 > \Delta \gamma$) described above. A large number of oscillations makes it possible to estimate the envelope. Its decay rate is only slightly lower than the PE decay rate without the field, so it is not possible to reliably extract the dark exciton dephasing rate $T_{2d}$ from their comparison. It also should be mentioned that at such a high magnetic field many effects should be taken into account including the diamagnetic shift of the exciton spectral position, nonlinearity and inhomogeneity of $g$-factors and possible magnetic field effects on dephasings. On the other hand the determination of $T_{2d}$ at lower fields is problematic since it will be hard to estimate the behaviour of the PE oscillations envelope.

Following the above consideration the long-living dark coherence could be revealed by the switching to the non-oscillatory regime with lowering the precession frequency. At the magnetic field below $1$~T such regime is observed. Fig.~\ref{Fig_Data}(b) shows PE decays for magnetic fields from 0 to 0.7~T. PE decays with field have a clear minimum at the temporal position shifting to shorter time delays with growing fields. After this minimum the slowly decaying tail is observed without the hint of second oscillation. This behaviour is consistent with the non-oscillatory regime realized at $\Omega_0 < \Delta \gamma$.

Another striking feature of the non-oscillatory regime could be observed in the magnetic field sweep experiment. The inset in Fig.~\ref{Fig_Data}(b) shows the magnetic field sweep of PE signal at $\tau = 250$~ps (denoted by asterisk in Fig.~~\ref{Fig_Data}(b)). This scan shows the global maximum of the PE signal at $B \approx 0.55$~T. We  want to stress that at such fields diamagnetic shift of the exciton transition is negligible, and it was checked that it plays no role in the formation of this maximum.

In order to examine the functional behaviour of the PE decay in the non-oscillatory regime we will focus on the PE signal we have observed in our experiment at $B = 0.4$~T (Fig.~\ref{Fig_Data}(c), red dots). To start with, this signal could be fitted by the equal dephasing rates model ($\Delta \gamma = 0$, green curve). Although this model could fit the initial decay and the first PE signal minimum, a second minimum is expected at $\tau = 270$~ps, which is obviously not observed in the experiment. The developed model with $\Delta \gamma \ne 0$ fits the data correctly yielding the $T_{2b} = 27 \pm 3$~ps and $T_{2d} = 240 \pm 30$~ps. The PE signal minimum around $100$~ps is not reaching pure zero as it is predicted by the theory. Its smoothing could be attributed to the non-zero exchange interaction and the $g$-factors distribution. These phenomena although are not affecting the non-oscillatory behaviour qualitatively, and will be studied in details and reported elsewhere.

% T2 = 1/(2*g), bright: g = g2+dg. dark: g = g2. (values from fit: g2 = 0.0021, dg = 0.016)

Long-living dark coherence could be used in coherence manipulation applications. We have shown that it could be accessed in the presence of the transverse magnetic field. The theoretical model for the PE in the case of different dephasing rates of states was developed, and the predicted non-oscillatory regime revealing dark coherence with the almost an order of magnitude longer decay time in comparison with the bright coherence was demonstrated in the experiment with heavy-hole excitons in the InGaAs/GaAs QW. This phenomenon could be used both for the study of coherent properties of dark states and as a protocol for optical coherence storage.

This study was supported by RFBR projects 19-52-12046 NNIO\_a and 20-32-70163. S.V.P. and I.A.Yu. thank the St. Petersburg State University (Grant No. 51125686). This work was carried out on the equipment of the SPbU Resource Center ''Nanophotonics''. We acknowledge Gleb.~G. Kozlov for fruitful discussions.

\bibliography{apssamp}% Produces the bibliography via BibTeX.

\providecommand{\noopsort}[1]{}\providecommand{\singleletter}[1]{#1}%
\begin{thebibliography}{40}%
\makeatletter
\providecommand \@ifxundefined [1]{%
 \@ifx{#1\undefined}
}%
\providecommand \@ifnum [1]{%
 \ifnum #1\expandafter \@firstoftwo
 \else \expandafter \@secondoftwo
 \fi
}%
\providecommand \@ifx [1]{%
 \ifx #1\expandafter \@firstoftwo
 \else \expandafter \@secondoftwo
 \fi
}%
\providecommand \natexlab [1]{#1}%
\providecommand \enquote  [1]{``#1''}%
\providecommand \bibnamefont  [1]{#1}%
\providecommand \bibfnamefont [1]{#1}%
\providecommand \citenamefont [1]{#1}%
\providecommand \href@noop [0]{\@secondoftwo}%
\providecommand \href [0]{\begingroup \@sanitize@url \@href}%
\providecommand \@href[1]{\@@startlink{#1}\@@href}%
\providecommand \@@href[1]{\endgroup#1\@@endlink}%
\providecommand \@sanitize@url [0]{\catcode `\\12\catcode `\$12\catcode
  `\&12\catcode `\#12\catcode `\^12\catcode `\_12\catcode `\%12\relax}%
\providecommand \@@startlink[1]{}%
\providecommand \@@endlink[0]{}%
\providecommand \url  [0]{\begingroup\@sanitize@url \@url }%
\providecommand \@url [1]{\endgroup\@href {#1}{\urlprefix }}%
\providecommand \urlprefix  [0]{URL }%
\providecommand \Eprint [0]{\href }%
\providecommand \doibase [0]{https://doi.org/}%
\providecommand \selectlanguage [0]{\@gobble}%
\providecommand \bibinfo  [0]{\@secondoftwo}%
\providecommand \bibfield  [0]{\@secondoftwo}%
\providecommand \translation [1]{[#1]}%
\providecommand \BibitemOpen [0]{}%
\providecommand \bibitemStop [0]{}%
\providecommand \bibitemNoStop [0]{.\EOS\space}%
\providecommand \EOS [0]{\spacefactor3000\relax}%
\providecommand \BibitemShut  [1]{\csname bibitem#1\endcsname}%
\let\auto@bib@innerbib\@empty
%</preamble>
\bibitem [{\citenamefont {Kim}\ \emph {et~al.}(2011)\citenamefont {Kim},
  \citenamefont {Carter}, \citenamefont {Greilich}, \citenamefont {Bracker},\
  and\ \citenamefont {Gammon}}]{Kim}%
  \BibitemOpen
  \bibfield  {author} {\bibinfo {author} {\bibfnamefont {D.}~\bibnamefont
  {Kim}}, \bibinfo {author} {\bibfnamefont {S.}~\bibnamefont {Carter}},
  \bibinfo {author} {\bibfnamefont {A.}~\bibnamefont {Greilich}}, \bibinfo
  {author} {\bibfnamefont {A.}~\bibnamefont {Bracker}},\ and\ \bibinfo {author}
  {\bibfnamefont {D.}~\bibnamefont {Gammon}},\ }\bibfield  {title} {\bibinfo
  {title} {Ultrafast optical control of entanglement between two quantum-dot
  spins},\ }\href {https://doi.org/10.1038/nphys1863} {\bibfield  {journal}
  {\bibinfo  {journal} {Nature Physics}\ }\textbf {\bibinfo {volume} {7}},\
  \bibinfo {pages} {223} (\bibinfo {year} {2011})}\BibitemShut {NoStop}%
\bibitem [{\citenamefont {Biolatti}\ \emph {et~al.}(2000)\citenamefont
  {Biolatti}, \citenamefont {Iotti}, \citenamefont {Zanardi},\ and\
  \citenamefont {Rossi}}]{Biolatti}%
  \BibitemOpen
  \bibfield  {author} {\bibinfo {author} {\bibfnamefont {E.}~\bibnamefont
  {Biolatti}}, \bibinfo {author} {\bibfnamefont {R.}~\bibnamefont {Iotti}},
  \bibinfo {author} {\bibfnamefont {P.}~\bibnamefont {Zanardi}},\ and\ \bibinfo
  {author} {\bibfnamefont {F.}~\bibnamefont {Rossi}},\ }\bibfield  {title}
  {\bibinfo {title} {Quantum information processing with semiconductor
  macroatoms},\ }\href {https://doi.org/10.1103/PhysRevLett.85.5647} {\bibfield
   {journal} {\bibinfo  {journal} {Physical Review Letters}\ }\textbf {\bibinfo
  {volume} {85}},\ \bibinfo {pages} {5647} (\bibinfo {year}
  {2000})}\BibitemShut {NoStop}%
\bibitem [{\citenamefont {Li}\ \emph {et~al.}(2003)\citenamefont {Li},
  \citenamefont {Wu}, \citenamefont {Steel}, \citenamefont {Gammon},
  \citenamefont {Stievater}, \citenamefont {Katzer}, \citenamefont {Park},
  \citenamefont {Piermarocchi},\ and\ \citenamefont {Sham}}]{Li}%
  \BibitemOpen
  \bibfield  {author} {\bibinfo {author} {\bibfnamefont {X.}~\bibnamefont
  {Li}}, \bibinfo {author} {\bibfnamefont {Y.}~\bibnamefont {Wu}}, \bibinfo
  {author} {\bibfnamefont {D.}~\bibnamefont {Steel}}, \bibinfo {author}
  {\bibfnamefont {D.}~\bibnamefont {Gammon}}, \bibinfo {author} {\bibfnamefont
  {T.}~\bibnamefont {Stievater}}, \bibinfo {author} {\bibfnamefont
  {D.}~\bibnamefont {Katzer}}, \bibinfo {author} {\bibfnamefont
  {D.}~\bibnamefont {Park}}, \bibinfo {author} {\bibfnamefont {C.}~\bibnamefont
  {Piermarocchi}},\ and\ \bibinfo {author} {\bibfnamefont {L.}~\bibnamefont
  {Sham}},\ }\bibfield  {title} {\bibinfo {title} {An all-optical quantum gate
  in a semiconductor quantum dot},\ }\href
  {https://doi.org/10.1126/science.1083800} {\bibfield  {journal} {\bibinfo
  {journal} {Science}\ }\textbf {\bibinfo {volume} {301}},\ \bibinfo {pages}
  {809} (\bibinfo {year} {2003})}\BibitemShut {NoStop}%
\bibitem [{\citenamefont {Kroutvar}\ \emph {et~al.}(2004)\citenamefont
  {Kroutvar}, \citenamefont {Ducommun}, \citenamefont {Heiss}, \citenamefont
  {Bichler}, \citenamefont {Schuh}, \citenamefont {Abstreiter},\ and\
  \citenamefont {Finley}}]{Kroutvar}%
  \BibitemOpen
  \bibfield  {author} {\bibinfo {author} {\bibfnamefont {M.}~\bibnamefont
  {Kroutvar}}, \bibinfo {author} {\bibfnamefont {Y.}~\bibnamefont {Ducommun}},
  \bibinfo {author} {\bibfnamefont {D.}~\bibnamefont {Heiss}}, \bibinfo
  {author} {\bibfnamefont {M.}~\bibnamefont {Bichler}}, \bibinfo {author}
  {\bibfnamefont {D.}~\bibnamefont {Schuh}}, \bibinfo {author} {\bibfnamefont
  {G.}~\bibnamefont {Abstreiter}},\ and\ \bibinfo {author} {\bibfnamefont
  {J.}~\bibnamefont {Finley}},\ }\bibfield  {title} {\bibinfo {title}
  {Optically programmable electron spin memory using semiconductor quantum
  dots},\ }\href {https://doi.org/10.1038/nature03008} {\bibfield  {journal}
  {\bibinfo  {journal} {Nature}\ }\textbf {\bibinfo {volume} {432}},\ \bibinfo
  {pages} {81} (\bibinfo {year} {2004})}\BibitemShut {NoStop}%
\bibitem [{\citenamefont {Henneberger}\ and\ \citenamefont
  {Benson}(2008)}]{Bits}%
  \BibitemOpen
  \bibfield  {author} {\bibinfo {author} {\bibfnamefont {F.}~\bibnamefont
  {Henneberger}}\ and\ \bibinfo {author} {\bibfnamefont {O.}~\bibnamefont
  {Benson}},\ }\href@noop {} {\emph {\bibinfo {title} {Semiconductor Quantum
  Bits}}}\ (\bibinfo  {publisher} {Jenny Stanford Publishing},\ \bibinfo {year}
  {2008})\BibitemShut {NoStop}%
\bibitem [{\citenamefont {Lvovsky}\ \emph {et~al.}(2009)\citenamefont
  {Lvovsky}, \citenamefont {Sanders},\ and\ \citenamefont {Tittel}}]{Lvovsky}%
  \BibitemOpen
  \bibfield  {author} {\bibinfo {author} {\bibfnamefont {A.}~\bibnamefont
  {Lvovsky}}, \bibinfo {author} {\bibfnamefont {B.}~\bibnamefont {Sanders}},\
  and\ \bibinfo {author} {\bibfnamefont {W.}~\bibnamefont {Tittel}},\
  }\bibfield  {title} {\bibinfo {title} {Optical quantum memory},\ }\href
  {https://doi.org/10.1038/nphoton.2009.231} {\bibfield  {journal} {\bibinfo
  {journal} {Nature Photonics}\ }\textbf {\bibinfo {volume} {3}},\ \bibinfo
  {pages} {706} (\bibinfo {year} {2009})}\BibitemShut {NoStop}%
\bibitem [{\citenamefont {Hammerer}\ \emph {et~al.}(2010)\citenamefont
  {Hammerer}, \citenamefont {Sørensen},\ and\ \citenamefont
  {Polzik}}]{Hammerer}%
  \BibitemOpen
  \bibfield  {author} {\bibinfo {author} {\bibfnamefont {K.}~\bibnamefont
  {Hammerer}}, \bibinfo {author} {\bibfnamefont {A.}~\bibnamefont
  {Sørensen}},\ and\ \bibinfo {author} {\bibfnamefont {E.}~\bibnamefont
  {Polzik}},\ }\bibfield  {title} {\bibinfo {title} {Quantum interface between
  light and atomic ensembles},\ }\href
  {https://doi.org/10.1103/RevModPhys.82.1041} {\bibfield  {journal} {\bibinfo
  {journal} {Reviews of Modern Physics}\ }\textbf {\bibinfo {volume} {82}},\
  \bibinfo {pages} {1041} (\bibinfo {year} {2010})}\BibitemShut {NoStop}%
\bibitem [{\citenamefont {Poltavtsev}\ \emph
  {et~al.}(2018{\natexlab{a}})\citenamefont {Poltavtsev}, \citenamefont
  {Yugova}, \citenamefont {Akimov}, \citenamefont {Yakovlev},\ and\
  \citenamefont {Bayer}}]{Polt2018PE}%
  \BibitemOpen
  \bibfield  {author} {\bibinfo {author} {\bibfnamefont {S.}~\bibnamefont
  {Poltavtsev}}, \bibinfo {author} {\bibfnamefont {I.}~\bibnamefont {Yugova}},
  \bibinfo {author} {\bibfnamefont {I.}~\bibnamefont {Akimov}}, \bibinfo
  {author} {\bibfnamefont {D.}~\bibnamefont {Yakovlev}},\ and\ \bibinfo
  {author} {\bibfnamefont {M.}~\bibnamefont {Bayer}},\ }\bibfield  {title}
  {\bibinfo {title} {Photon echo from localized excitons in semiconductor
  nanostructures},\ }\href {https://doi.org/10.1134/S1063783418080188}
  {\bibfield  {journal} {\bibinfo  {journal} {Physics of the Solid State}\
  }\textbf {\bibinfo {volume} {60}},\ \bibinfo {pages} {1635} (\bibinfo {year}
  {2018}{\natexlab{a}})}\BibitemShut {NoStop}%
\bibitem [{\citenamefont {Solovev}\ \emph
  {et~al.}(2018{\natexlab{a}})\citenamefont {Solovev}, \citenamefont
  {Kapitonov}, \citenamefont {Stroganov}, \citenamefont {Efimov}, \citenamefont
  {Eliseev},\ and\ \citenamefont {Poltavtsev}}]{Sol2018InGaAs}%
  \BibitemOpen
  \bibfield  {author} {\bibinfo {author} {\bibfnamefont {I.}~\bibnamefont
  {Solovev}}, \bibinfo {author} {\bibfnamefont {Y.}~\bibnamefont {Kapitonov}},
  \bibinfo {author} {\bibfnamefont {B.}~\bibnamefont {Stroganov}}, \bibinfo
  {author} {\bibfnamefont {Y.}~\bibnamefont {Efimov}}, \bibinfo {author}
  {\bibfnamefont {S.}~\bibnamefont {Eliseev}},\ and\ \bibinfo {author}
  {\bibfnamefont {S.}~\bibnamefont {Poltavtsev}},\ }\bibfield  {title}
  {\bibinfo {title} {Separation of inhomogeneous and homogeneous broadening
  manifestations in {InGaAs/GaAs} quantum wells by time-resolved four-wave
  mixing}\ }(\bibinfo {year} {2018})\BibitemShut {NoStop}%
\bibitem [{\citenamefont {Salewski}\ \emph
  {et~al.}(2017{\natexlab{a}})\citenamefont {Salewski}, \citenamefont
  {Poltavtsev}, \citenamefont {Kapitonov}, \citenamefont {Vondran},
  \citenamefont {Yakovlev}, \citenamefont {Schneider}, \citenamefont {Kamp},
  \citenamefont {Höfling}, \citenamefont {Oulton}, \citenamefont {Akimov},
  \citenamefont {Kavokin},\ and\ \citenamefont {Bayer}}]{Sal2017InGaAsQD}%
  \BibitemOpen
  \bibfield  {author} {\bibinfo {author} {\bibfnamefont {M.}~\bibnamefont
  {Salewski}}, \bibinfo {author} {\bibfnamefont {S.}~\bibnamefont
  {Poltavtsev}}, \bibinfo {author} {\bibfnamefont {Y.}~\bibnamefont
  {Kapitonov}}, \bibinfo {author} {\bibfnamefont {J.}~\bibnamefont {Vondran}},
  \bibinfo {author} {\bibfnamefont {D.}~\bibnamefont {Yakovlev}}, \bibinfo
  {author} {\bibfnamefont {C.}~\bibnamefont {Schneider}}, \bibinfo {author}
  {\bibfnamefont {M.}~\bibnamefont {Kamp}}, \bibinfo {author} {\bibfnamefont
  {S.}~\bibnamefont {Höfling}}, \bibinfo {author} {\bibfnamefont
  {R.}~\bibnamefont {Oulton}}, \bibinfo {author} {\bibfnamefont
  {I.}~\bibnamefont {Akimov}}, \bibinfo {author} {\bibfnamefont
  {A.}~\bibnamefont {Kavokin}},\ and\ \bibinfo {author} {\bibfnamefont
  {M.}~\bibnamefont {Bayer}},\ }\bibfield  {title} {\bibinfo {title} {Photon
  echoes from {(In,Ga)As} quantum dots embedded in a {Tamm-plasmon}
  microcavity},\ }\bibfield  {journal} {\bibinfo  {journal} {Physical Review
  B}\ }\textbf {\bibinfo {volume} {95}},\ \href
  {https://doi.org/10.1103/PhysRevB.95.035312} {10.1103/PhysRevB.95.035312}
  (\bibinfo {year} {2017}{\natexlab{a}})\BibitemShut {NoStop}%
\bibitem [{\citenamefont {Poltavtsev}\ \emph {et~al.}(2016)\citenamefont
  {Poltavtsev}, \citenamefont {Salewski}, \citenamefont {Kapitonov},
  \citenamefont {Yugova}, \citenamefont {Akimov}, \citenamefont {Schneider},
  \citenamefont {Kamp}, \citenamefont {Höfling}, \citenamefont {Yakovlev},
  \citenamefont {Kavokin},\ and\ \citenamefont {Bayer}}]{Polt2016InGaAsQD}%
  \BibitemOpen
  \bibfield  {author} {\bibinfo {author} {\bibfnamefont {S.}~\bibnamefont
  {Poltavtsev}}, \bibinfo {author} {\bibfnamefont {M.}~\bibnamefont
  {Salewski}}, \bibinfo {author} {\bibfnamefont {Y.}~\bibnamefont {Kapitonov}},
  \bibinfo {author} {\bibfnamefont {I.}~\bibnamefont {Yugova}}, \bibinfo
  {author} {\bibfnamefont {I.}~\bibnamefont {Akimov}}, \bibinfo {author}
  {\bibfnamefont {C.}~\bibnamefont {Schneider}}, \bibinfo {author}
  {\bibfnamefont {M.}~\bibnamefont {Kamp}}, \bibinfo {author} {\bibfnamefont
  {S.}~\bibnamefont {Höfling}}, \bibinfo {author} {\bibfnamefont
  {D.}~\bibnamefont {Yakovlev}}, \bibinfo {author} {\bibfnamefont
  {A.}~\bibnamefont {Kavokin}},\ and\ \bibinfo {author} {\bibfnamefont
  {M.}~\bibnamefont {Bayer}},\ }\bibfield  {title} {\bibinfo {title} {Photon
  echo transients from an inhomogeneous ensemble of semiconductor quantum
  dots},\ }\bibfield  {journal} {\bibinfo  {journal} {Physical Review B}\
  }\textbf {\bibinfo {volume} {93}},\ \href
  {https://doi.org/10.1103/PhysRevB.93.121304} {10.1103/PhysRevB.93.121304}
  (\bibinfo {year} {2016})\BibitemShut {NoStop}%
\bibitem [{\citenamefont {Poltavtsev}\ \emph
  {et~al.}(2017{\natexlab{a}})\citenamefont {Poltavtsev}, \citenamefont
  {Reichelt}, \citenamefont {Akimov}, \citenamefont {Karczewski}, \citenamefont
  {Wiater}, \citenamefont {Wojtowicz}, \citenamefont {Yakovlev}, \citenamefont
  {Meier},\ and\ \citenamefont {Bayer}}]{Polt2017Rabi}%
  \BibitemOpen
  \bibfield  {author} {\bibinfo {author} {\bibfnamefont {S.}~\bibnamefont
  {Poltavtsev}}, \bibinfo {author} {\bibfnamefont {M.}~\bibnamefont
  {Reichelt}}, \bibinfo {author} {\bibfnamefont {I.}~\bibnamefont {Akimov}},
  \bibinfo {author} {\bibfnamefont {G.}~\bibnamefont {Karczewski}}, \bibinfo
  {author} {\bibfnamefont {M.}~\bibnamefont {Wiater}}, \bibinfo {author}
  {\bibfnamefont {T.}~\bibnamefont {Wojtowicz}}, \bibinfo {author}
  {\bibfnamefont {D.}~\bibnamefont {Yakovlev}}, \bibinfo {author}
  {\bibfnamefont {T.}~\bibnamefont {Meier}},\ and\ \bibinfo {author}
  {\bibfnamefont {M.}~\bibnamefont {Bayer}},\ }\bibfield  {title} {\bibinfo
  {title} {Damping of {Rabi} oscillations in intensity-dependent photon echoes
  from exciton complexes in a {CdTe/(Cd,Mg)Te} single quantum well},\
  }\bibfield  {journal} {\bibinfo  {journal} {Physical Review B}\ }\textbf
  {\bibinfo {volume} {96}},\ \href {https://doi.org/10.1103/PhysRevB.96.075306}
  {10.1103/PhysRevB.96.075306} (\bibinfo {year}
  {2017}{\natexlab{a}})\BibitemShut {NoStop}%
\bibitem [{\citenamefont {Salewski}\ \emph
  {et~al.}(2017{\natexlab{b}})\citenamefont {Salewski}, \citenamefont
  {Poltavtsev}, \citenamefont {Yugova}, \citenamefont {Karczewski},
  \citenamefont {Wiater}, \citenamefont {Wojtowicz}, \citenamefont {Yakovlev},
  \citenamefont {Akimov}, \citenamefont {Meier},\ and\ \citenamefont
  {Bayer}}]{Sal2017PRX}%
  \BibitemOpen
  \bibfield  {author} {\bibinfo {author} {\bibfnamefont {M.}~\bibnamefont
  {Salewski}}, \bibinfo {author} {\bibfnamefont {S.}~\bibnamefont
  {Poltavtsev}}, \bibinfo {author} {\bibfnamefont {I.}~\bibnamefont {Yugova}},
  \bibinfo {author} {\bibfnamefont {G.}~\bibnamefont {Karczewski}}, \bibinfo
  {author} {\bibfnamefont {M.}~\bibnamefont {Wiater}}, \bibinfo {author}
  {\bibfnamefont {T.}~\bibnamefont {Wojtowicz}}, \bibinfo {author}
  {\bibfnamefont {D.}~\bibnamefont {Yakovlev}}, \bibinfo {author}
  {\bibfnamefont {I.}~\bibnamefont {Akimov}}, \bibinfo {author} {\bibfnamefont
  {T.}~\bibnamefont {Meier}},\ and\ \bibinfo {author} {\bibfnamefont
  {M.}~\bibnamefont {Bayer}},\ }\bibfield  {title} {\bibinfo {title}
  {High-resolution two-dimensional optical spectroscopy of electron spins},\
  }\bibfield  {journal} {\bibinfo  {journal} {Physical Review X}\ }\textbf
  {\bibinfo {volume} {7}},\ \href {https://doi.org/10.1103/PhysRevX.7.031030}
  {10.1103/PhysRevX.7.031030} (\bibinfo {year}
  {2017}{\natexlab{b}})\BibitemShut {NoStop}%
\bibitem [{\citenamefont {Langer}\ \emph {et~al.}(2014)\citenamefont {Langer},
  \citenamefont {Poltavtsev}, \citenamefont {Yugova}, \citenamefont {Salewski},
  \citenamefont {Yakovlev}, \citenamefont {Karczewski}, \citenamefont
  {Wojtowicz}, \citenamefont {Akimov},\ and\ \citenamefont
  {Bayer}}]{Lang2014NatPh}%
  \BibitemOpen
  \bibfield  {author} {\bibinfo {author} {\bibfnamefont {L.}~\bibnamefont
  {Langer}}, \bibinfo {author} {\bibfnamefont {S.}~\bibnamefont {Poltavtsev}},
  \bibinfo {author} {\bibfnamefont {I.}~\bibnamefont {Yugova}}, \bibinfo
  {author} {\bibfnamefont {M.}~\bibnamefont {Salewski}}, \bibinfo {author}
  {\bibfnamefont {D.}~\bibnamefont {Yakovlev}}, \bibinfo {author}
  {\bibfnamefont {G.}~\bibnamefont {Karczewski}}, \bibinfo {author}
  {\bibfnamefont {T.}~\bibnamefont {Wojtowicz}}, \bibinfo {author}
  {\bibfnamefont {I.}~\bibnamefont {Akimov}},\ and\ \bibinfo {author}
  {\bibfnamefont {M.}~\bibnamefont {Bayer}},\ }\bibfield  {title} {\bibinfo
  {title} {Access to long-term optical memories using photon echoes retrieved
  from semiconductor spins},\ }\href {https://doi.org/10.1038/nphoton.2014.219}
  {\bibfield  {journal} {\bibinfo  {journal} {Nature Photonics}\ }\textbf
  {\bibinfo {volume} {8}},\ \bibinfo {pages} {851} (\bibinfo {year}
  {2014})}\BibitemShut {NoStop}%
\bibitem [{\citenamefont {Langer}\ \emph {et~al.}(2012)\citenamefont {Langer},
  \citenamefont {Poltavtsev}, \citenamefont {Yugova}, \citenamefont {Yakovlev},
  \citenamefont {Karczewski}, \citenamefont {Wojtowicz}, \citenamefont
  {Kossut}, \citenamefont {Akimov},\ and\ \citenamefont {Bayer}}]{Lang2012PRL}%
  \BibitemOpen
  \bibfield  {author} {\bibinfo {author} {\bibfnamefont {L.}~\bibnamefont
  {Langer}}, \bibinfo {author} {\bibfnamefont {S.}~\bibnamefont {Poltavtsev}},
  \bibinfo {author} {\bibfnamefont {I.}~\bibnamefont {Yugova}}, \bibinfo
  {author} {\bibfnamefont {D.}~\bibnamefont {Yakovlev}}, \bibinfo {author}
  {\bibfnamefont {G.}~\bibnamefont {Karczewski}}, \bibinfo {author}
  {\bibfnamefont {T.}~\bibnamefont {Wojtowicz}}, \bibinfo {author}
  {\bibfnamefont {J.}~\bibnamefont {Kossut}}, \bibinfo {author} {\bibfnamefont
  {I.}~\bibnamefont {Akimov}},\ and\ \bibinfo {author} {\bibfnamefont
  {M.}~\bibnamefont {Bayer}},\ }\bibfield  {title} {\bibinfo {title}
  {Magnetic-field control of photon echo from the electron-trion system in a
  {CdTe} quantum well: Shuffling coherence between optically accessible and
  inaccessible states},\ }\bibfield  {journal} {\bibinfo  {journal} {Physical
  Review Letters}\ }\textbf {\bibinfo {volume} {109}},\ \href
  {https://doi.org/10.1103/PhysRevLett.109.157403}
  {10.1103/PhysRevLett.109.157403} (\bibinfo {year} {2012})\BibitemShut
  {NoStop}%
\bibitem [{\citenamefont {Solovev}\ \emph
  {et~al.}(2018{\natexlab{b}})\citenamefont {Solovev}, \citenamefont
  {Poltavtsev}, \citenamefont {Kapitonov}, \citenamefont {Akimov},
  \citenamefont {Sadofev}, \citenamefont {Puls}, \citenamefont {Yakovlev},\
  and\ \citenamefont {Bayer}}]{Sol2018ZnO}%
  \BibitemOpen
  \bibfield  {author} {\bibinfo {author} {\bibfnamefont {I.}~\bibnamefont
  {Solovev}}, \bibinfo {author} {\bibfnamefont {S.}~\bibnamefont {Poltavtsev}},
  \bibinfo {author} {\bibfnamefont {Y.}~\bibnamefont {Kapitonov}}, \bibinfo
  {author} {\bibfnamefont {I.}~\bibnamefont {Akimov}}, \bibinfo {author}
  {\bibfnamefont {S.}~\bibnamefont {Sadofev}}, \bibinfo {author} {\bibfnamefont
  {J.}~\bibnamefont {Puls}}, \bibinfo {author} {\bibfnamefont {D.}~\bibnamefont
  {Yakovlev}},\ and\ \bibinfo {author} {\bibfnamefont {M.}~\bibnamefont
  {Bayer}},\ }\bibfield  {title} {\bibinfo {title} {Coherent dynamics of
  localized excitons and trions in {ZnO/(Zn,Mg)O} quantum wells studied by
  photon echoes},\ }\bibfield  {journal} {\bibinfo  {journal} {Physical Review
  B}\ }\textbf {\bibinfo {volume} {97}},\ \href
  {https://doi.org/10.1103/PhysRevB.97.245406} {10.1103/PhysRevB.97.245406}
  (\bibinfo {year} {2018}{\natexlab{b}})\BibitemShut {NoStop}%
\bibitem [{\citenamefont {Poltavtsev}\ \emph
  {et~al.}(2017{\natexlab{b}})\citenamefont {Poltavtsev}, \citenamefont
  {Kosarev}, \citenamefont {Akimov}, \citenamefont {Yakovlev}, \citenamefont
  {Sadofev}, \citenamefont {Puls}, \citenamefont {Hoffmann}, \citenamefont
  {Albert}, \citenamefont {Meier}, \citenamefont {Meier},\ and\ \citenamefont
  {Bayer}}]{Polt2017ZnO}%
  \BibitemOpen
  \bibfield  {author} {\bibinfo {author} {\bibfnamefont {S.}~\bibnamefont
  {Poltavtsev}}, \bibinfo {author} {\bibfnamefont {A.}~\bibnamefont {Kosarev}},
  \bibinfo {author} {\bibfnamefont {I.}~\bibnamefont {Akimov}}, \bibinfo
  {author} {\bibfnamefont {D.}~\bibnamefont {Yakovlev}}, \bibinfo {author}
  {\bibfnamefont {S.}~\bibnamefont {Sadofev}}, \bibinfo {author} {\bibfnamefont
  {J.}~\bibnamefont {Puls}}, \bibinfo {author} {\bibfnamefont {S.}~\bibnamefont
  {Hoffmann}}, \bibinfo {author} {\bibfnamefont {M.}~\bibnamefont {Albert}},
  \bibinfo {author} {\bibfnamefont {C.}~\bibnamefont {Meier}}, \bibinfo
  {author} {\bibfnamefont {T.}~\bibnamefont {Meier}},\ and\ \bibinfo {author}
  {\bibfnamefont {M.}~\bibnamefont {Bayer}},\ }\bibfield  {title} {\bibinfo
  {title} {Time-resolved photon echoes from donor-bound excitons in {ZnO}
  epitaxial layers},\ }\bibfield  {journal} {\bibinfo  {journal} {Physical
  Review B}\ }\textbf {\bibinfo {volume} {96}},\ \href
  {https://doi.org/10.1103/PhysRevB.96.035203} {10.1103/PhysRevB.96.035203}
  (\bibinfo {year} {2017}{\natexlab{b}})\BibitemShut {NoStop}%
\bibitem [{\citenamefont {Poltavtsev}\ \emph
  {et~al.}(2018{\natexlab{b}})\citenamefont {Poltavtsev}, \citenamefont
  {Solovev}, \citenamefont {Akimov}, \citenamefont {Chaldyshev}, \citenamefont
  {Lundin}, \citenamefont {Sakharov}, \citenamefont {Tsatsulnikov},
  \citenamefont {Yakovlev},\ and\ \citenamefont {Bayer}}]{Polt2018GaN}%
  \BibitemOpen
  \bibfield  {author} {\bibinfo {author} {\bibfnamefont {S.}~\bibnamefont
  {Poltavtsev}}, \bibinfo {author} {\bibfnamefont {I.}~\bibnamefont {Solovev}},
  \bibinfo {author} {\bibfnamefont {I.}~\bibnamefont {Akimov}}, \bibinfo
  {author} {\bibfnamefont {V.}~\bibnamefont {Chaldyshev}}, \bibinfo {author}
  {\bibfnamefont {W.}~\bibnamefont {Lundin}}, \bibinfo {author} {\bibfnamefont
  {A.}~\bibnamefont {Sakharov}}, \bibinfo {author} {\bibfnamefont
  {A.}~\bibnamefont {Tsatsulnikov}}, \bibinfo {author} {\bibfnamefont
  {D.}~\bibnamefont {Yakovlev}},\ and\ \bibinfo {author} {\bibfnamefont
  {M.}~\bibnamefont {Bayer}},\ }\bibfield  {title} {\bibinfo {title} {Long
  coherent dynamics of localized excitons in {(In,Ga)N/GaN} quantum wells},\
  }\bibfield  {journal} {\bibinfo  {journal} {Physical Review B}\ }\textbf
  {\bibinfo {volume} {98}},\ \href {https://doi.org/10.1103/PhysRevB.98.195315}
  {10.1103/PhysRevB.98.195315} (\bibinfo {year}
  {2018}{\natexlab{b}})\BibitemShut {NoStop}%
\bibitem [{\citenamefont {Hao}\ \emph {et~al.}(2016)\citenamefont {Hao},
  \citenamefont {Moody}, \citenamefont {Wu}, \citenamefont {Dass},
  \citenamefont {Xu}, \citenamefont {Chen}, \citenamefont {Sun}, \citenamefont
  {Li}, \citenamefont {Li}, \citenamefont {MacDonald},\ and\ \citenamefont
  {Li}}]{TMDC2016}%
  \BibitemOpen
  \bibfield  {author} {\bibinfo {author} {\bibfnamefont {K.}~\bibnamefont
  {Hao}}, \bibinfo {author} {\bibfnamefont {G.}~\bibnamefont {Moody}}, \bibinfo
  {author} {\bibfnamefont {F.}~\bibnamefont {Wu}}, \bibinfo {author}
  {\bibfnamefont {C.}~\bibnamefont {Dass}}, \bibinfo {author} {\bibfnamefont
  {L.}~\bibnamefont {Xu}}, \bibinfo {author} {\bibfnamefont {C.-H.}\
  \bibnamefont {Chen}}, \bibinfo {author} {\bibfnamefont {L.}~\bibnamefont
  {Sun}}, \bibinfo {author} {\bibfnamefont {M.-Y.}\ \bibnamefont {Li}},
  \bibinfo {author} {\bibfnamefont {L.-J.}\ \bibnamefont {Li}}, \bibinfo
  {author} {\bibfnamefont {A.}~\bibnamefont {MacDonald}},\ and\ \bibinfo
  {author} {\bibfnamefont {X.}~\bibnamefont {Li}},\ }\bibfield  {title}
  {\bibinfo {title} {Direct measurement of exciton valley coherence in
  monolayer {WSe2}},\ }\href {https://doi.org/10.1038/nphys3674} {\bibfield
  {journal} {\bibinfo  {journal} {Nature Physics}\ }\textbf {\bibinfo {volume}
  {12}},\ \bibinfo {pages} {677} (\bibinfo {year} {2016})}\BibitemShut
  {NoStop}%
\bibitem [{\citenamefont {Jakubczyk}\ \emph {et~al.}(2016)\citenamefont
  {Jakubczyk}, \citenamefont {Delmonte}, \citenamefont {Koperski},
  \citenamefont {Nogajewski}, \citenamefont {Faugeras}, \citenamefont
  {Langbein}, \citenamefont {Potemski},\ and\ \citenamefont
  {Kasprzak}}]{KaspTMDC2016}%
  \BibitemOpen
  \bibfield  {author} {\bibinfo {author} {\bibfnamefont {T.}~\bibnamefont
  {Jakubczyk}}, \bibinfo {author} {\bibfnamefont {V.}~\bibnamefont {Delmonte}},
  \bibinfo {author} {\bibfnamefont {M.}~\bibnamefont {Koperski}}, \bibinfo
  {author} {\bibfnamefont {K.}~\bibnamefont {Nogajewski}}, \bibinfo {author}
  {\bibfnamefont {C.}~\bibnamefont {Faugeras}}, \bibinfo {author}
  {\bibfnamefont {W.}~\bibnamefont {Langbein}}, \bibinfo {author}
  {\bibfnamefont {M.}~\bibnamefont {Potemski}},\ and\ \bibinfo {author}
  {\bibfnamefont {J.}~\bibnamefont {Kasprzak}},\ }\bibfield  {title} {\bibinfo
  {title} {Radiatively limited dephasing and exciton dynamics in {MoSe2}
  monolayers revealed with four-wave mixing microscopy},\ }\href
  {https://doi.org/10.1021/acs.nanolett.6b01060} {\bibfield  {journal}
  {\bibinfo  {journal} {Nano Letters}\ }\textbf {\bibinfo {volume} {16}},\
  \bibinfo {pages} {5333} (\bibinfo {year} {2016})}\BibitemShut {NoStop}%
\bibitem [{\citenamefont {March}\ \emph {et~al.}(2016)\citenamefont {March},
  \citenamefont {Clegg}, \citenamefont {Riley}, \citenamefont {Webber},
  \citenamefont {Hill},\ and\ \citenamefont {Hall}}]{March2016}%
  \BibitemOpen
  \bibfield  {author} {\bibinfo {author} {\bibfnamefont {S.}~\bibnamefont
  {March}}, \bibinfo {author} {\bibfnamefont {C.}~\bibnamefont {Clegg}},
  \bibinfo {author} {\bibfnamefont {D.}~\bibnamefont {Riley}}, \bibinfo
  {author} {\bibfnamefont {D.}~\bibnamefont {Webber}}, \bibinfo {author}
  {\bibfnamefont {I.}~\bibnamefont {Hill}},\ and\ \bibinfo {author}
  {\bibfnamefont {K.}~\bibnamefont {Hall}},\ }\bibfield  {title} {\bibinfo
  {title} {Simultaneous observation of free and defect-bound excitons in
  {CH3NH3PbI3} using four-wave mixing spectroscopy},\ }\bibfield  {journal}
  {\bibinfo  {journal} {Scientific Reports}\ }\textbf {\bibinfo {volume} {6}},\
  \href {https://doi.org/10.1038/srep39139} {10.1038/srep39139} (\bibinfo
  {year} {2016})\BibitemShut {NoStop}%
\bibitem [{\citenamefont {Bohn}\ \emph {et~al.}(2018)\citenamefont {Bohn},
  \citenamefont {Simon}, \citenamefont {Gramlich}, \citenamefont {Richter},
  \citenamefont {Polavarapu}, \citenamefont {Urban},\ and\ \citenamefont
  {Feldmann}}]{Bohn2018}%
  \BibitemOpen
  \bibfield  {author} {\bibinfo {author} {\bibfnamefont {B.}~\bibnamefont
  {Bohn}}, \bibinfo {author} {\bibfnamefont {T.}~\bibnamefont {Simon}},
  \bibinfo {author} {\bibfnamefont {M.}~\bibnamefont {Gramlich}}, \bibinfo
  {author} {\bibfnamefont {A.}~\bibnamefont {Richter}}, \bibinfo {author}
  {\bibfnamefont {L.}~\bibnamefont {Polavarapu}}, \bibinfo {author}
  {\bibfnamefont {A.}~\bibnamefont {Urban}},\ and\ \bibinfo {author}
  {\bibfnamefont {J.}~\bibnamefont {Feldmann}},\ }\bibfield  {title} {\bibinfo
  {title} {Dephasing and quantum beating of excitons in methylammonium lead
  iodide perovskite nanoplatelets},\ }\href
  {https://doi.org/10.1021/acsphotonics.7b01292} {\bibfield  {journal}
  {\bibinfo  {journal} {ACS Photonics}\ }\textbf {\bibinfo {volume} {5}},\
  \bibinfo {pages} {648} (\bibinfo {year} {2018})}\BibitemShut {NoStop}%
\bibitem [{\citenamefont {March}\ \emph {et~al.}(2017)\citenamefont {March},
  \citenamefont {Riley}, \citenamefont {Clegg}, \citenamefont {Webber},
  \citenamefont {Todd}, \citenamefont {Hill},\ and\ \citenamefont
  {Hall}}]{March2017}%
  \BibitemOpen
  \bibfield  {author} {\bibinfo {author} {\bibfnamefont {S.}~\bibnamefont
  {March}}, \bibinfo {author} {\bibfnamefont {D.}~\bibnamefont {Riley}},
  \bibinfo {author} {\bibfnamefont {C.}~\bibnamefont {Clegg}}, \bibinfo
  {author} {\bibfnamefont {D.}~\bibnamefont {Webber}}, \bibinfo {author}
  {\bibfnamefont {S.}~\bibnamefont {Todd}}, \bibinfo {author} {\bibfnamefont
  {I.}~\bibnamefont {Hill}},\ and\ \bibinfo {author} {\bibfnamefont
  {K.}~\bibnamefont {Hall}},\ }\bibfield  {title} {\bibinfo {title} {Four-wave
  mixing response of solution-processed {CH3NH3PbI3} thin films}\ }(\bibinfo
  {year} {2017})\BibitemShut {NoStop}%
\bibitem [{\citenamefont {March}\ \emph {et~al.}(2019)\citenamefont {March},
  \citenamefont {Riley}, \citenamefont {Clegg}, \citenamefont {Webber},
  \citenamefont {Hill}, \citenamefont {Yu},\ and\ \citenamefont
  {Hall}}]{March2019}%
  \BibitemOpen
  \bibfield  {author} {\bibinfo {author} {\bibfnamefont {S.}~\bibnamefont
  {March}}, \bibinfo {author} {\bibfnamefont {D.}~\bibnamefont {Riley}},
  \bibinfo {author} {\bibfnamefont {C.}~\bibnamefont {Clegg}}, \bibinfo
  {author} {\bibfnamefont {D.}~\bibnamefont {Webber}}, \bibinfo {author}
  {\bibfnamefont {I.}~\bibnamefont {Hill}}, \bibinfo {author} {\bibfnamefont
  {Z.-G.}\ \bibnamefont {Yu}},\ and\ \bibinfo {author} {\bibfnamefont
  {K.}~\bibnamefont {Hall}},\ }\bibfield  {title} {\bibinfo {title} {Ultrafast
  acoustic phonon scattering in {CH3NH3PbI3} revealed by femtosecond four-wave
  mixing},\ }\bibfield  {journal} {\bibinfo  {journal} {Journal of Chemical
  Physics}\ }\textbf {\bibinfo {volume} {151}},\ \href
  {https://doi.org/10.1063/1.5120385} {10.1063/1.5120385} (\bibinfo {year}
  {2019})\BibitemShut {NoStop}%
\bibitem [{\citenamefont {Poltavtsev}\ \emph {et~al.}(2020)\citenamefont
  {Poltavtsev}, \citenamefont {Yugova}, \citenamefont {Babenko}, \citenamefont
  {Akimov}, \citenamefont {Yakovlev}, \citenamefont {Karczewski}, \citenamefont
  {Chusnutdinow}, \citenamefont {Wojtowicz},\ and\ \citenamefont
  {Bayer}}]{Polt2020PRB}%
  \BibitemOpen
  \bibfield  {author} {\bibinfo {author} {\bibfnamefont {S.~V.}\ \bibnamefont
  {Poltavtsev}}, \bibinfo {author} {\bibfnamefont {I.~A.}\ \bibnamefont
  {Yugova}}, \bibinfo {author} {\bibfnamefont {I.~A.}\ \bibnamefont {Babenko}},
  \bibinfo {author} {\bibfnamefont {I.~A.}\ \bibnamefont {Akimov}}, \bibinfo
  {author} {\bibfnamefont {D.~R.}\ \bibnamefont {Yakovlev}}, \bibinfo {author}
  {\bibfnamefont {G.}~\bibnamefont {Karczewski}}, \bibinfo {author}
  {\bibfnamefont {S.}~\bibnamefont {Chusnutdinow}}, \bibinfo {author}
  {\bibfnamefont {T.}~\bibnamefont {Wojtowicz}},\ and\ \bibinfo {author}
  {\bibfnamefont {M.}~\bibnamefont {Bayer}},\ }\bibfield  {title} {\bibinfo
  {title} {Quantum beats in the polarization of the spin-dependent photon echo
  from donor-bound excitons in {CdTe/(Cd,Mg)Te} quantum wells},\ }\href
  {https://doi.org/10.1103/PhysRevB.101.081409} {\bibfield  {journal} {\bibinfo
   {journal} {Phys. Rev. B}\ }\textbf {\bibinfo {volume} {101}},\ \bibinfo
  {pages} {081409} (\bibinfo {year} {2020})}\BibitemShut {NoStop}%
\bibitem [{\citenamefont {Poltavtsev}\ \emph {et~al.}(2019)\citenamefont
  {Poltavtsev}, \citenamefont {Kapitonov}, \citenamefont {Yugova},
  \citenamefont {Akimov}, \citenamefont {Yakovlev}, \citenamefont {Karczewski},
  \citenamefont {Wiater}, \citenamefont {Wojtowicz},\ and\ \citenamefont
  {Bayer}}]{Polt2019Sci}%
  \BibitemOpen
  \bibfield  {author} {\bibinfo {author} {\bibfnamefont {S.}~\bibnamefont
  {Poltavtsev}}, \bibinfo {author} {\bibfnamefont {Y.}~\bibnamefont
  {Kapitonov}}, \bibinfo {author} {\bibfnamefont {I.}~\bibnamefont {Yugova}},
  \bibinfo {author} {\bibfnamefont {I.}~\bibnamefont {Akimov}}, \bibinfo
  {author} {\bibfnamefont {D.}~\bibnamefont {Yakovlev}}, \bibinfo {author}
  {\bibfnamefont {G.}~\bibnamefont {Karczewski}}, \bibinfo {author}
  {\bibfnamefont {M.}~\bibnamefont {Wiater}}, \bibinfo {author} {\bibfnamefont
  {T.}~\bibnamefont {Wojtowicz}},\ and\ \bibinfo {author} {\bibfnamefont
  {M.}~\bibnamefont {Bayer}},\ }\bibfield  {title} {\bibinfo {title}
  {Polarimetry of photon echo on charged and neutral excitons in semiconductor
  quantum wells},\ }\href@noop {} {\bibfield  {journal} {\bibinfo  {journal}
  {Scientific Reports}\ }\textbf {\bibinfo {volume} {9}} (\bibinfo {year}
  {2019})}\BibitemShut {NoStop}%
\bibitem [{\citenamefont {Glasberg}\ \emph {et~al.}(1999)\citenamefont
  {Glasberg}, \citenamefont {Shtrikman}, \citenamefont {Bar-Joseph},\ and\
  \citenamefont {Klipstein}}]{Glasberg1999}%
  \BibitemOpen
  \bibfield  {author} {\bibinfo {author} {\bibfnamefont {S.}~\bibnamefont
  {Glasberg}}, \bibinfo {author} {\bibfnamefont {H.}~\bibnamefont {Shtrikman}},
  \bibinfo {author} {\bibfnamefont {I.}~\bibnamefont {Bar-Joseph}},\ and\
  \bibinfo {author} {\bibfnamefont {P.~C.}\ \bibnamefont {Klipstein}},\
  }\bibfield  {title} {\bibinfo {title} {{Exciton exchange splitting in wide
  GaAs quantum wells}},\ }\href {https://doi.org/10.1103/PhysRevB.60.R16295}
  {\bibfield  {journal} {\bibinfo  {journal} {Phys. Rev. B}\ }\textbf {\bibinfo
  {volume} {60}},\ \bibinfo {pages} {R16295} (\bibinfo {year}
  {1999})}\BibitemShut {NoStop}%
\bibitem [{\citenamefont {Bayer}\ \emph {et~al.}(2000)\citenamefont {Bayer},
  \citenamefont {Stern}, \citenamefont {Kuther},\ and\ \citenamefont
  {Forchel}}]{Bayer2000}%
  \BibitemOpen
  \bibfield  {author} {\bibinfo {author} {\bibfnamefont {M.}~\bibnamefont
  {Bayer}}, \bibinfo {author} {\bibfnamefont {O.}~\bibnamefont {Stern}},
  \bibinfo {author} {\bibfnamefont {A.}~\bibnamefont {Kuther}},\ and\ \bibinfo
  {author} {\bibfnamefont {A.}~\bibnamefont {Forchel}},\ }\bibfield  {title}
  {\bibinfo {title} {{Spectroscopic study of dark excitons in
  ${\mathrm{In}}_{x}{\mathrm{Ga}}_{1\ensuremath{-}x}\mathrm{As}$ self-assembled
  quantum dots by a magnetic-field-induced symmetry breaking}},\ }\href
  {https://doi.org/10.1103/PhysRevB.61.7273} {\bibfield  {journal} {\bibinfo
  {journal} {Phys. Rev. B}\ }\textbf {\bibinfo {volume} {61}},\ \bibinfo
  {pages} {7273} (\bibinfo {year} {2000})}\BibitemShut {NoStop}%
\bibitem [{\citenamefont {Zaric}(2004)}]{Zaric2004}%
  \BibitemOpen
  \bibfield  {author} {\bibinfo {author} {\bibfnamefont {S.}~\bibnamefont
  {Zaric}},\ }\bibfield  {title} {\bibinfo {title} {{Optical Signatures of the
  Aharonov-Bohm Phase in Single-Walled Carbon Nanotubes}},\ }\href
  {https://doi.org/10.1126/science.1096524} {\bibfield  {journal} {\bibinfo
  {journal} {Science (80-. ).}\ }\textbf {\bibinfo {volume} {304}},\ \bibinfo
  {pages} {1129} (\bibinfo {year} {2004})}\BibitemShut {NoStop}%
\bibitem [{\citenamefont {Lu}\ \emph {et~al.}(2020)\citenamefont {Lu},
  \citenamefont {Rhodes}, \citenamefont {Li}, \citenamefont {van Tuan},
  \citenamefont {Jiang}, \citenamefont {Ludwig}, \citenamefont {Jiang},
  \citenamefont {Lian}, \citenamefont {Shi}, \citenamefont {Hone},
  \citenamefont {Dery},\ and\ \citenamefont {Smirnov}}]{Lu2020}%
  \BibitemOpen
  \bibfield  {author} {\bibinfo {author} {\bibfnamefont {Z.}~\bibnamefont
  {Lu}}, \bibinfo {author} {\bibfnamefont {D.}~\bibnamefont {Rhodes}}, \bibinfo
  {author} {\bibfnamefont {Z.}~\bibnamefont {Li}}, \bibinfo {author}
  {\bibfnamefont {D.}~\bibnamefont {van Tuan}}, \bibinfo {author}
  {\bibfnamefont {Y.}~\bibnamefont {Jiang}}, \bibinfo {author} {\bibfnamefont
  {J.}~\bibnamefont {Ludwig}}, \bibinfo {author} {\bibfnamefont
  {Z.}~\bibnamefont {Jiang}}, \bibinfo {author} {\bibfnamefont
  {Z.}~\bibnamefont {Lian}}, \bibinfo {author} {\bibfnamefont {S.~F.}\
  \bibnamefont {Shi}}, \bibinfo {author} {\bibfnamefont {J.}~\bibnamefont
  {Hone}}, \bibinfo {author} {\bibfnamefont {H.}~\bibnamefont {Dery}},\ and\
  \bibinfo {author} {\bibfnamefont {D.}~\bibnamefont {Smirnov}},\ }\bibfield
  {title} {\bibinfo {title} {{Magnetic field mixing and splitting of bright and
  dark excitons in monolayer MoSe2}},\ }\bibfield  {journal} {\bibinfo
  {journal} {2D Mater.}\ }\textbf {\bibinfo {volume} {7}},\ \href
  {https://doi.org/10.1088/2053-1583/AB5614} {10.1088/2053-1583/AB5614}
  (\bibinfo {year} {2020}),\ \Eprint {https://arxiv.org/abs/1905.10439}
  {arXiv:1905.10439} \BibitemShut {NoStop}%
\bibitem [{\citenamefont {Zhang}\ \emph {et~al.}(2017)\citenamefont {Zhang},
  \citenamefont {Cao}, \citenamefont {Lu}, \citenamefont {Lin}, \citenamefont
  {Zhang}, \citenamefont {Wang}, \citenamefont {Li}, \citenamefont {Hone},
  \citenamefont {Robinson}, \citenamefont {Smirnov}, \citenamefont {Louie},\
  and\ \citenamefont {Heinz}}]{Zhang2017}%
  \BibitemOpen
  \bibfield  {author} {\bibinfo {author} {\bibfnamefont {X.~X.}\ \bibnamefont
  {Zhang}}, \bibinfo {author} {\bibfnamefont {T.}~\bibnamefont {Cao}}, \bibinfo
  {author} {\bibfnamefont {Z.}~\bibnamefont {Lu}}, \bibinfo {author}
  {\bibfnamefont {Y.~C.}\ \bibnamefont {Lin}}, \bibinfo {author} {\bibfnamefont
  {F.}~\bibnamefont {Zhang}}, \bibinfo {author} {\bibfnamefont
  {Y.}~\bibnamefont {Wang}}, \bibinfo {author} {\bibfnamefont {Z.}~\bibnamefont
  {Li}}, \bibinfo {author} {\bibfnamefont {J.~C.}\ \bibnamefont {Hone}},
  \bibinfo {author} {\bibfnamefont {J.~A.}\ \bibnamefont {Robinson}}, \bibinfo
  {author} {\bibfnamefont {D.}~\bibnamefont {Smirnov}}, \bibinfo {author}
  {\bibfnamefont {S.~G.}\ \bibnamefont {Louie}},\ and\ \bibinfo {author}
  {\bibfnamefont {T.~F.}\ \bibnamefont {Heinz}},\ }\bibfield  {title} {\bibinfo
  {title} {{Magnetic brightening and control of dark excitons in monolayer WSe
  2}},\ }\href {https://doi.org/10.1038/nnano.2017.105} {\bibfield  {journal}
  {\bibinfo  {journal} {Nat. Nanotechnol.}\ }\textbf {\bibinfo {volume} {12}},\
  \bibinfo {pages} {883} (\bibinfo {year} {2017})}\BibitemShut {NoStop}%
\bibitem [{\citenamefont {Tang}\ \emph {et~al.}(2019)\citenamefont {Tang},
  \citenamefont {Mak},\ and\ \citenamefont {Shan}}]{Tang2019}%
  \BibitemOpen
  \bibfield  {author} {\bibinfo {author} {\bibfnamefont {Y.}~\bibnamefont
  {Tang}}, \bibinfo {author} {\bibfnamefont {K.~F.}\ \bibnamefont {Mak}},\ and\
  \bibinfo {author} {\bibfnamefont {J.}~\bibnamefont {Shan}},\ }\bibfield
  {title} {\bibinfo {title} {{Long valley lifetime of dark excitons in
  single-layer WSe2}},\ }\href {https://doi.org/10.1038/s41467-019-12129-1}
  {\bibfield  {journal} {\bibinfo  {journal} {Nat. Commun.}\ }\textbf {\bibinfo
  {volume} {10}},\ \bibinfo {pages} {4047} (\bibinfo {year}
  {2019})}\BibitemShut {NoStop}%
\bibitem [{\citenamefont {Zhou}\ \emph {et~al.}(2017)\citenamefont {Zhou},
  \citenamefont {Scuri}, \citenamefont {Wild}, \citenamefont {High},
  \citenamefont {Dibos}, \citenamefont {Jauregui}, \citenamefont {Shu},
  \citenamefont {{De Greve}}, \citenamefont {Pistunova}, \citenamefont {Joe},
  \citenamefont {Taniguchi}, \citenamefont {Watanabe}, \citenamefont {Kim},
  \citenamefont {Lukin},\ and\ \citenamefont {Park}}]{Zhou2017}%
  \BibitemOpen
  \bibfield  {author} {\bibinfo {author} {\bibfnamefont {Y.}~\bibnamefont
  {Zhou}}, \bibinfo {author} {\bibfnamefont {G.}~\bibnamefont {Scuri}},
  \bibinfo {author} {\bibfnamefont {D.~S.}\ \bibnamefont {Wild}}, \bibinfo
  {author} {\bibfnamefont {A.~A.}\ \bibnamefont {High}}, \bibinfo {author}
  {\bibfnamefont {A.}~\bibnamefont {Dibos}}, \bibinfo {author} {\bibfnamefont
  {L.~A.}\ \bibnamefont {Jauregui}}, \bibinfo {author} {\bibfnamefont
  {C.}~\bibnamefont {Shu}}, \bibinfo {author} {\bibfnamefont {K.}~\bibnamefont
  {{De Greve}}}, \bibinfo {author} {\bibfnamefont {K.}~\bibnamefont
  {Pistunova}}, \bibinfo {author} {\bibfnamefont {A.~Y.}\ \bibnamefont {Joe}},
  \bibinfo {author} {\bibfnamefont {T.}~\bibnamefont {Taniguchi}}, \bibinfo
  {author} {\bibfnamefont {K.}~\bibnamefont {Watanabe}}, \bibinfo {author}
  {\bibfnamefont {P.}~\bibnamefont {Kim}}, \bibinfo {author} {\bibfnamefont
  {M.~D.}\ \bibnamefont {Lukin}},\ and\ \bibinfo {author} {\bibfnamefont
  {H.}~\bibnamefont {Park}},\ }\bibfield  {title} {\bibinfo {title} {{Probing
  dark excitons in atomically thin semiconductors via near-field coupling to
  surface plasmon polaritons}},\ }\href
  {https://doi.org/10.1038/nnano.2017.106} {\bibfield  {journal} {\bibinfo
  {journal} {Nat. Nanotechnol.}\ }\textbf {\bibinfo {volume} {12}},\ \bibinfo
  {pages} {856} (\bibinfo {year} {2017})},\ \Eprint
  {https://arxiv.org/abs/1701.05938} {arXiv:1701.05938} \BibitemShut {NoStop}%
\bibitem [{\citenamefont {Ikeuchi}\ \emph {et~al.}(2003)\citenamefont
  {Ikeuchi}, \citenamefont {Adachi}, \citenamefont {Sasakura},\ and\
  \citenamefont {Muto}}]{Ikeuchi2003}%
  \BibitemOpen
  \bibfield  {author} {\bibinfo {author} {\bibfnamefont {O.}~\bibnamefont
  {Ikeuchi}}, \bibinfo {author} {\bibfnamefont {S.}~\bibnamefont {Adachi}},
  \bibinfo {author} {\bibfnamefont {H.}~\bibnamefont {Sasakura}},\ and\
  \bibinfo {author} {\bibfnamefont {S.}~\bibnamefont {Muto}},\ }\bibfield
  {title} {\bibinfo {title} {{Observation of population transfer to dark
  exciton states by using spin-diffracted four-wave mixing}},\ }\href
  {https://doi.org/10.1063/1.1575920} {\bibfield  {journal} {\bibinfo
  {journal} {J. Appl. Phys.}\ }\textbf {\bibinfo {volume} {93}},\ \bibinfo
  {pages} {9634} (\bibinfo {year} {2003})}\BibitemShut {NoStop}%
\bibitem [{\citenamefont {Tomoda}\ \emph {et~al.}(2010)\citenamefont {Tomoda},
  \citenamefont {Adachi}, \citenamefont {Muto},\ and\ \citenamefont
  {Shimomura}}]{Tomoda2010}%
  \BibitemOpen
  \bibfield  {author} {\bibinfo {author} {\bibfnamefont {K.}~\bibnamefont
  {Tomoda}}, \bibinfo {author} {\bibfnamefont {S.}~\bibnamefont {Adachi}},
  \bibinfo {author} {\bibfnamefont {S.}~\bibnamefont {Muto}},\ and\ \bibinfo
  {author} {\bibfnamefont {S.}~\bibnamefont {Shimomura}},\ }\bibfield  {title}
  {\bibinfo {title} {{Transient grating studies of phase and spin relaxations
  of excitons in GaAs single quantum wells}},\ }\href
  {https://doi.org/10.1016/j.physe.2009.12.009} {\bibfield  {journal} {\bibinfo
   {journal} {Phys. E Low-dimensional Syst. Nanostructures}\ }\textbf {\bibinfo
  {volume} {42}},\ \bibinfo {pages} {2714} (\bibinfo {year}
  {2010})}\BibitemShut {NoStop}%
\bibitem [{SMA()}]{SMA}%
  \BibitemOpen
  \href@noop {} {}\bibinfo {note} {See Supplemental Material at [URL] for the
  details of the PE calculation.}\BibitemShut {Stop}%
\bibitem [{SMB()}]{SMB}%
  \BibitemOpen
  \href@noop {} {}\bibinfo {note} {See Supplemental Material at [URL] for the
  description of the mechanical analogy.}\BibitemShut {Stop}%
\bibitem [{\citenamefont {Poltavtsev}\ \emph {et~al.}(2014)\citenamefont
  {Poltavtsev}, \citenamefont {Efimov}, \citenamefont {Dolgikh}, \citenamefont
  {Eliseev}, \citenamefont {Petrov},\ and\ \citenamefont
  {Ovsyankin}}]{Polt2014Brew}%
  \BibitemOpen
  \bibfield  {author} {\bibinfo {author} {\bibfnamefont {S.}~\bibnamefont
  {Poltavtsev}}, \bibinfo {author} {\bibfnamefont {Y.}~\bibnamefont {Efimov}},
  \bibinfo {author} {\bibfnamefont {Y.}~\bibnamefont {Dolgikh}}, \bibinfo
  {author} {\bibfnamefont {S.}~\bibnamefont {Eliseev}}, \bibinfo {author}
  {\bibfnamefont {V.}~\bibnamefont {Petrov}},\ and\ \bibinfo {author}
  {\bibfnamefont {V.}~\bibnamefont {Ovsyankin}},\ }\bibfield  {title} {\bibinfo
  {title} {Extremely low inhomogeneous broadening of exciton lines in shallow
  {(In,Ga)As/GaAs} quantum wells},\ }\href
  {https://doi.org/10.1016/j.ssc.2014.09.005} {\bibfield  {journal} {\bibinfo
  {journal} {Solid State Communications}\ }\textbf {\bibinfo {volume} {199}},\
  \bibinfo {pages} {47} (\bibinfo {year} {2014})}\BibitemShut {NoStop}%
\bibitem [{\citenamefont {Poltavtsev}\ and\ \citenamefont
  {Stroganov}(2010)}]{Polt2010Brew}%
  \BibitemOpen
  \bibfield  {author} {\bibinfo {author} {\bibfnamefont {S.}~\bibnamefont
  {Poltavtsev}}\ and\ \bibinfo {author} {\bibfnamefont {B.}~\bibnamefont
  {Stroganov}},\ }\bibfield  {title} {\bibinfo {title} {Experimental
  investigation of the oscillator strength of the exciton transition in {GaAs}
  single quantum wells},\ }\href {https://doi.org/10.1134/S1063783410090180}
  {\bibfield  {journal} {\bibinfo  {journal} {Physics of the Solid State}\
  }\textbf {\bibinfo {volume} {52}},\ \bibinfo {pages} {1899} (\bibinfo {year}
  {2010})}\BibitemShut {NoStop}%
\bibitem [{SMC()}]{SMC}%
  \BibitemOpen
  \href@noop {} {}\bibinfo {note} {See Supplemental Material at [URL] for the
  exciton representation in the linear polarizations basis.}\BibitemShut
  {Stop}%
\end{thebibliography}%

\end{document}

\section{Supplementary materials}

\subsection{Photon echo from the three-level system}

\begin{figure}[h]
\includegraphics[width=0.3 \linewidth]{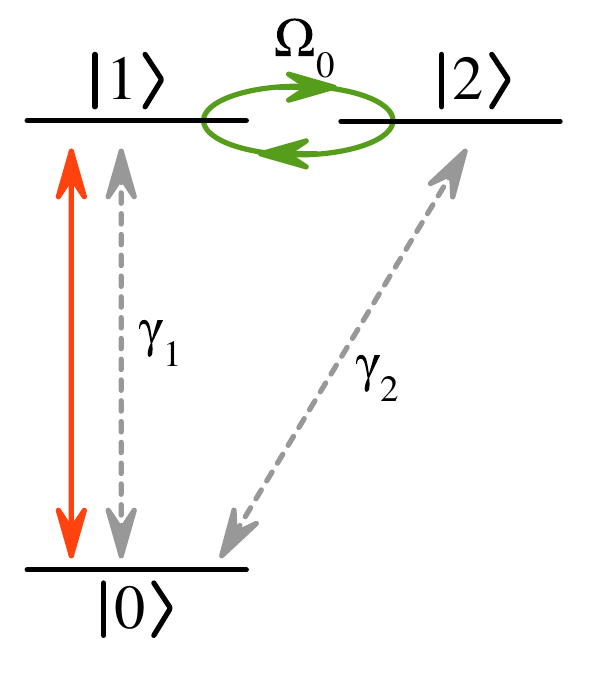}
\caption{\label{FigSMScheme} Energy diagram of the three-level system with the optically addressed transition $|0\rangle \leftrightarrow |1\rangle$ (red arrow) and different dephasing rates for $|0\rangle \leftrightarrow |1\rangle$ and $|0\rangle \leftrightarrow |2\rangle$ polarizations (gray arrows). States $|1\rangle$ and $|2\rangle$ are coupled with the precession frequency $\Omega_0$.}
\end{figure}

Let us consider the three-level system depicted in Fig.~\ref{FigSMScheme}. The ground state $|0\rangle$ and the excited state $|1\rangle$ are coupled by the light. States $|1\rangle$ and $|2\rangle$ are coupled with the precession frequency $\Omega_0$. This system could be described by the density matrix $\rho$, where $\rho_{11}$ and $\rho_{22}$ are the matrix elements representing populations of excited states $|1\rangle$ and $|2\rangle$ correspondingly, and $\rho_{01} = \rho_{10}^*$ and $\rho_{02} = \rho_{20}^*$ are matrix elements responsible for optical polarizations.

The temporal evolution of the system could be found using the Lindblad equation:

\begin{equation}
    \label{Lind}
    i \dot{\rho} = -\frac{i}{\hbar} [\hat{H}, \rho] + \Gamma,
\end{equation}{}

where $\Gamma$ is the phenomenologically introduced decays matrix.

The Hamiltonian describing the interaction with light and temporal evolution of the system could be written in the following form:

\begin{equation}
    \hat{H} =
    \left(
    \begin{array}{ccc}
        0                   & \mathsf d^* E(t)^* e^{i \omega t}  & 0  \\
        \mathsf d E(t)e^{-i \omega t}   & \hbar\omega_0          & \hbar\Omega_0\\
        0                   & \hbar\Omega_0          & \hbar\omega_0
    \end{array}
    \right),
\end{equation}

where $E(t)$ is the envelope of the electric field in the pulse, $\mathsf d$ is the dipole moment of an optical transition, $\omega_0$ is the transition frequency, and $\omega$ is the light frequency. 

We will use the infinitely-short pulses approximation (Hahn echo regime). In this case the spontaneous photon echo experiment could be broken down into the sequence of actions of light pulses with areas $\Theta = \lim_{t_p \to \infty} \left( \frac{2}{\hbar} | \mathsf d E(t) t_p \right |)$, and temporal evolutions of the density matrix in the absence of the light field. The initial state of the system has only single non-zero density matrix element $\rho_{00} = 1$. 

For clarity we consider the simple spontaneous PE sequence with the first pulse area $\Theta_1 = \frac{\pi}{2}$ and the second pulse area $\Theta_2 = \pi$ separated by the time delay $\tau$. After the first $\frac{\pi}{2}$-pulse action $\rho_{01}(0) = -\frac{i}{2}$, $\rho_{02}(0) =0$. The action of the second $\pi$-pulse leads to the following changes of element of interest: $\rho_{01}(\tau+0) = \rho_{01}(\tau)^*+C_1$, $\rho_{02}(\tau+0) =C_2$, where $\rho_{0j}(\tau)$ and $\rho_{0j}(\tau+0)$ denotes the matrix elements before and after the pulse action correspondingly, and $C_1$ and $C_2$ are terms that do not contribute in the formation of the PE signal. 

The element $\rho_{01}(\tau)$ is determined by the temporal evolution between pulses. We consider this problem in more detail in the rotating-wave approximation (RWA), denoting the density matrix in RWA as $\tilde{\rho}$. The temporal evolution is represented by phenomenological decays, and the action of the Hamiltonian $\hat{H}_T$: 

\begin{equation}
    \hat{H}_T =
    \left(
    \begin{array}{ccc}
        0                   & 0  & 0  \\
        0   & 0          & \hbar\Omega_0\\
        0                   & \hbar\Omega_0         & 0
    \end{array}
    \right).
\end{equation}

We will consider different phenomenological dephasing rates $\gamma_1$ and $\gamma_2$ of optical polarizations $|0\rangle \leftrightarrow |1\rangle$ and $|0\rangle \leftrightarrow |2\rangle$ correspondingly (we will assume $\gamma_2 \le \gamma_1$). Temporal evolution and decays actions combined lead to the following set of differential equations:

\begin{equation}
    \label{time}
    \left\{
        \begin{array}{l}
            i \dot{\tilde{\rho}}_{01} =  - \Omega_0 \tilde{\rho}_{02} - i \gamma_1 \tilde{\rho}_{01} \\
            i \dot{\tilde{\rho}}_{02} = - \Omega_0 \tilde{\rho}_{01} - i \gamma_2 \tilde{\rho}_{02}             
        \end{array}
    \right..
\end{equation}

Differential equations for $\tilde{\rho}_{01}$ and $\tilde{\rho}_{02}$ are decoupled from equations for other density matrix elements. This system could be simplified if we rewrite it in the form of optical Bloch equations with elements $u_\alpha = 2 {\rm Re} (\tilde{\rho}_{0\alpha})$, $v_\alpha = -2 {\rm Im} (\tilde{\rho}_{0\alpha})$ ($\alpha = 1, 2$) by noting that $u_1 = v_2 = 0$ through the whole experiment:

\begin{equation}
    \label{bloch}
    \left\{
        \begin{array}{l}
        \dot{v}_{1} = - \Omega_0 u_{2} - \gamma_1 v_{1} \\
        \dot{u}_{2} =   \phantom{-} \Omega_0 v_{1} - \gamma_2 u_{2} 
        \end{array}
    \right..
\end{equation}

The first $\frac{\pi}{2}$-pulse action gives the following initial conditions for the above system: $v_1(0) =1$ and $u_2(0) = 0$. The solution could be found for the evolution during the time $t=\tau$ after the pulse action:

\begin{equation}
    \label{uvsol}
    \left\{
    \begin{array}{l}
        v_1(\tau) = e^{-\gamma \tau} \left( \cos (\Omega \tau) 
        - \frac{\Delta \gamma}{\Omega} \sin (\Omega \tau) \right)\\
        u_2(\tau) = \vphantom{\left( \frac{\Delta \gamma}{\Omega} \right)} e^{-\gamma \tau} \frac{\Omega_0}{\Omega} \sin (\Omega \tau)
    \end{array}
    \right.,
\end{equation}

where $\gamma = \frac{\gamma_1 + \gamma_2}{2}$ is the average dephasing rate, $\Delta \gamma = \frac{\gamma_1 - \gamma_2}{2}$ is the dephasing rates difference, and $\Omega = \sqrt{\Omega_0^2 - \Delta \gamma^2}$ is the generalized Larmor frequency. This evolution also could be represented in polar coordinates $r = \sqrt{v_1^2 + u_2^2}$ and $\alpha = \arctan \frac{u_2}{v_1}$:

\begin{equation}
    \left\{
    \begin{array}{l}
        r(\tau) = e^{-\gamma \tau} 
        \sqrt{1 - \frac{2 \Delta \gamma}{\Omega} \sin(\Omega \tau) \cos(\Omega \tau) + \frac{2 \Delta \gamma^2}{\Omega^2} \sin^2 (\Omega \tau) }
        \\
        \alpha(\tau) =
        \arctan \left(
            \frac{ \tan( \Omega \tau) }{ \frac{\Omega}{\Omega_0} - \frac{\Delta \gamma}{\Omega_0} \tan (\Omega \tau) }
        \right)
    \end{array}
    \right..
\end{equation}

The polarization $P$ of the system could be found as
$P = {\rm Tr}(\hat{d}\rho)= 2 {\rm Re} \left( \mathsf d \rho_{01}(\omega_0,t) \right)$, here $\hat{d}$ is the dipole moment operator. 
To calculate the PE signal from the ensemble one has to sum $P(\omega_0,t)$ over all $\omega_0$. 
Finally, the amplitude of the PE signal at $t=2\tau$ is the following:

\begin{equation}
        P_{PE} \sim 
        e^{- 2\gamma \tau} \left| \cos (\Omega \tau) - \frac{\Delta \gamma}{\Omega} \sin (\Omega \tau) \right|^2.
\end{equation}

The set of zeros for the $P_{PE}$ could be tracked back to zeroes of $v_1(\tau)$ in Eq.~(\ref{uvsol}). Depending on the sign of the expression under the radical in $\Omega$ the set of zeroes $\tau_k$ could represent oscillatory ($\Omega_0 > \Delta \gamma$), critical ($\Omega_0 = \Delta \gamma$) and non-oscillatory ($\Omega_0 < \Delta \gamma$) regimes of the polarization shuffling:

\begin{equation}
\tau_k =
\left\{
\begin{array}{cl}
	\frac{\frac{\pi}{2} - \arcsin \left( \frac{\Delta \gamma}{\Omega_0} \right) + \pi k}{\sqrt{\Omega_0^2 - \Delta \gamma^2}},
	&
	k \in Z, \, \Omega_0 > \Delta \gamma  
\\
\mathstrut
\\
	\frac{1}{\Delta \gamma},
	&
	k = 0, \, \Omega_0 = \Delta \gamma
\\
\mathstrut
\\
	\frac{\ln \left(
	    \frac{
	        \Delta \gamma + \sqrt{\Delta \gamma^2 - \Omega_0^2}}{\Omega_0} 
	   \right)}{\sqrt{\Delta \gamma^2 - \Omega_0^2}}
	,
	&
	k = 0, \, \Omega_0 < \Delta \gamma
\\
\end{array}
\right.
,
\end{equation}

PE signal could be found explicitly for these three regimes:

\begin{equation}
\label{eq_p}
P_{PE} \sim
\left\{
\begin{array}{ll}
        e^{- 2\gamma \tau}
        \left(
            \cos (\Omega \tau)
            -
            \frac{\Delta \gamma}{\Omega}
            \sin (\Omega \tau)
        \right)^2,
	&
	\Omega_0 > \Delta \gamma;
\\
\mathstrut
\\
	e^{-2 \gamma \tau}
	(1 - \Delta \gamma \tau )^2,
	&
	\Omega_0 = \Delta \gamma;
\\
\mathstrut
\\
        e^{- 2\gamma \tau}
        \left(
            \cosh (|\Omega | \tau)
            -
            \frac{\Delta \gamma}{|\Omega|}
            \sinh (|\Omega | \tau)
        \right)^2,
	&
	\Omega_0 < \Delta \gamma.
\\
\end{array}
\right.
\end{equation}

\begin{figure}
\includegraphics[width=0.8 \linewidth]{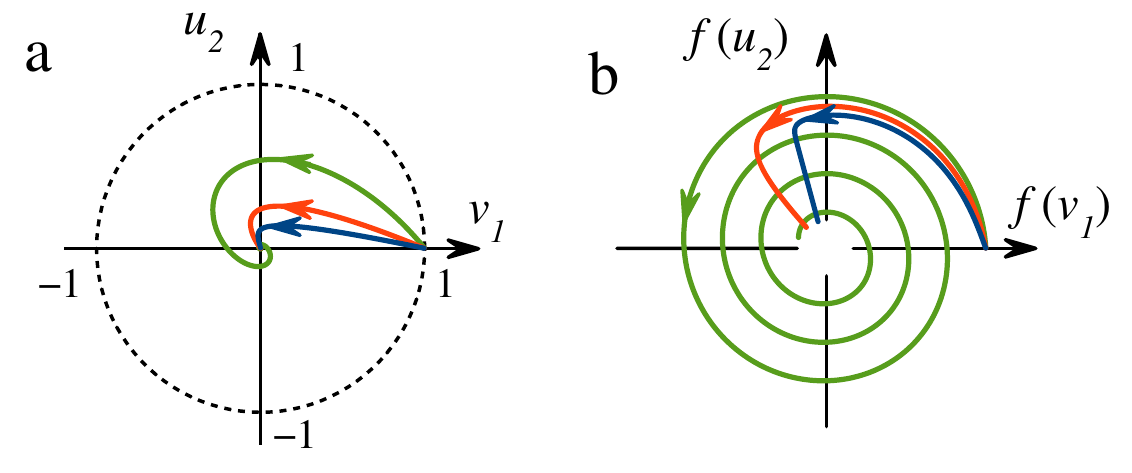}
\caption{\label{FigSpir} (a) Phase portraits for oscillatory ($\Omega_0 = 3 \Delta \gamma$, green curve), critical ($\Omega_0 = \Delta \gamma$, red curve) and non-oscillatory ($\Omega_0 = 0.5 \Delta \gamma$, blue curve) regimes. $\Delta \gamma / \gamma = 0.7$. 
(b) Same portraits shown on the  $(f(v_{1}),f(u_{2}))$-plane, where $f(x) = \frac{x}{r} (\log(r)+c)$ represents the log-polar coordinates transformation, $r = \sqrt{v^2_{1} + u^2_{2}}$ and offset $c$ is selected for visualisation clarity.}
\end{figure}

These three regimes could be represented on the $(v_1,u_2)$ phase plane (Fig.~\ref{FigSpir}(a)) as a phase portrait circling in a spiral around the central point for the oscillatory regime (green curve), and phase portraits infinitely ''falling'' on central point in critical (red curve) and non-oscillatory (blue curve) regimes. Figure~\ref{FigSpir}(b) shows the same curves in log-polar coordinates emphasising the difference between regimes.

The signal in the oscillatory regime ($\Omega_0 > \Delta \gamma$) could be represented in the following form with the introduction of $\sin \beta = \frac{\Delta \gamma}{\Omega_0}$:

\begin{equation}
P_{PE} \sim
	e^{-2 \gamma \tau}
	\frac{\cos^2
	(
		\Omega_0 \tau \cos \beta
		+
		\beta
	)}
	{
	    \cos^2 \beta
	},
	\,
	\Omega_0 > \Delta \gamma.
\end{equation}

This representation underline the analogy between oscillating cases with the dephasings difference ($\Delta \gamma \ne 0$) and without it ($\Delta \gamma = 0$) .

The signal in the non-oscillatory regime ($\Omega_0 < \Delta \gamma$) could be represented in the following form:

\begin{equation}
P_{PE} \sim
    \frac{\alpha^2}{(1-\alpha)^2}
    e^{ -2 (\gamma_2 + \alpha \Delta \gamma) \tau}
    +
    \frac{\alpha(2-\alpha)}{(1-\alpha)^2}
    e^{ -2 (\gamma_2 + \Delta \gamma) \tau}
    +
    \frac{(2-\alpha)^2}{(1-\alpha)^2}
    e^{ -2 (\gamma_2 + (2-\alpha) \Delta \gamma) \tau},
\end{equation}

where $\alpha = 1 - \sqrt{1 - \frac{\Omega_0^2}{\Delta \gamma^2}}$ (the non-oscillatory regime corresponds to $0 \le \alpha < 1$). It could be seen that at relatively big $\tau$ only the first term plays role, and the PE signal decays monoexponentially with the rate $2 (\gamma_2 + \alpha \Delta \gamma)$, which is the closer to dark coherence dephasing rate $2\gamma_2$, the less is $\Omega_0$ (and, consequently, the less is $\alpha$).

\subsection{Mechanical analogy}

\begin{figure}[h]
\includegraphics[width=0.4 \linewidth]{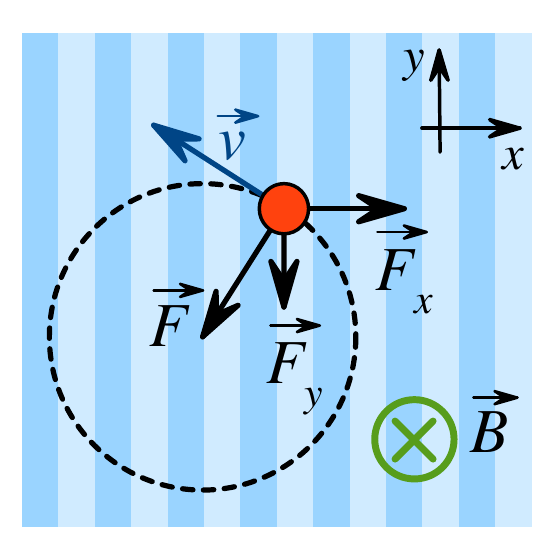}
\caption{\label{FigMech} Mechanical analogy to the problem: a charged particle (red) is moving in the anisotropic medium with the magnetic field $\vec{B}$ applied.}
\end{figure}

Lets us find a mechanical analogy to the problem studied. Consider a charged particle with mass $m$ and charge $q>0$ moving through the medium with the anisotropic viscous friction described by two coefficients $\eta_x$ and $\eta_y$ (Fig.~\ref{FigMech}). The magnetic field $\vec{B}$ is applied normal to the $(x,y)$ plane.

The Newton's second law for this particle reads as follows: $ m \dot{\vec{v}} = \vec{F} + \vec{F}_x + \vec{F}_y$, where $\vec{v}$ is the particle speed, the Lorentz force is $\vec{F} = q [ \vec{v} \times \vec{B} ]$ and the anisotropic friction force has two components: $F_x = -\eta_x v_x$ and $F_y = -\eta_y v_y$. In this situation the componentwise system reads:

\begin{equation}
    \left\{
        \begin{array}{l}
            m \dot{v}_x = - q B v_y - \eta_x v_x \\
            m \dot{v}_y=          \phantom{-}  q B v_x - \eta_y v_y
        \end{array}
    \right..
\end{equation}

With the substitution $v_x \to v_1$, $v_y \to u_2$, $\Omega_0 \to \frac{q}{m} B$, $\gamma_{1,2} \to \frac{\eta_{x,y}}{m}$ this system is equivalent to the Eq.~\ref{bloch}.

%\textcolor{red}{What is the physical meaning of the equivalence of polarization in PE and velocity in mechanics? Maybe we should find the system not for velocity but for coordinates x(t), y(t)? If so, then how to do it?}

\subsection{Exciton in the linear polarizations basis}

\begin{figure}
\includegraphics[width=0.8 \linewidth]{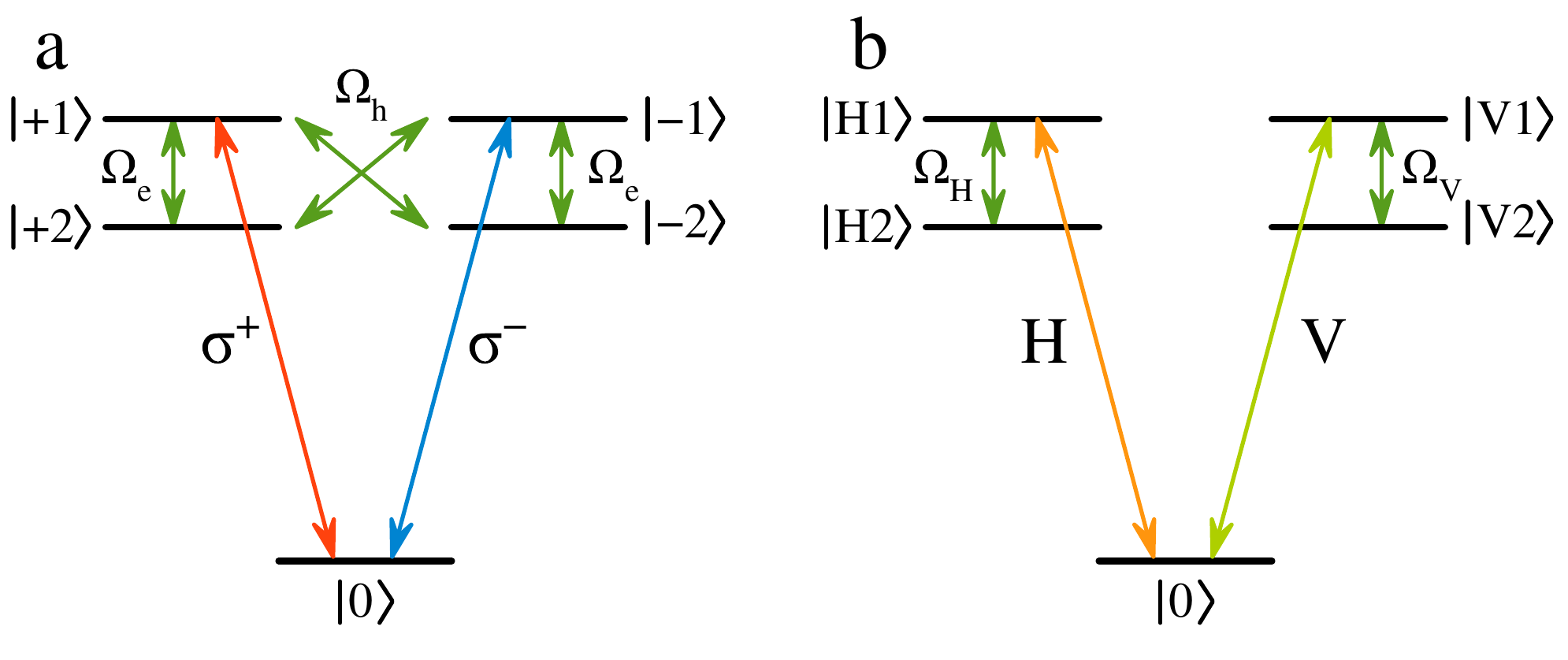}
\caption{\label{FigBases} Energy diagram of the exciton in circular (a) and linear (b) bases. The energy splitting of excited levels is introduced for clarity.}
\end{figure}

The energy diagram of the heavy-hole exciton in QW in the circular polarizations basis consist of the ground state $|0\rangle$, two bright exciton states $|\pm1 \rangle$ and two dark exciton states $|\pm 2\rangle$ (Fig.~\ref{FigBases}(a)). These states are coupled by the Larmor precession of the electron and hole spins in transverse magnetic field with frequencies $\Omega_e$ and $\Omega_h$ correspondingly. For simplicity, we neglect the exchange interaction and consider the isotropic transverse g-factors of the electron and the hole. The $\sigma^+$ ($\sigma^-$) polarized light with amplitude $E_+$ ($E_-$) couples $|+1\rangle$ ($|-1\rangle$) state with the $|0\rangle$ state. The light frequency is denoted as $\omega$, the system frequency is $\omega_0$ and circular components of the transition dipole moment are $d_{\pm}$.

In the circular polarizations basis $(|0\rangle, |+1\rangle, |-1\rangle, |+2\rangle, |-2\rangle)$ the $5\times5$ Hamiltonian has the following form:

\begin{equation}
    \hat{H}_{circ} =
    \left(
    \begin{array}{ccccc}
        0                               & d_+^*E_+^*e^{i \omega t} & d_-^*E_-^* e^{i \omega t}                 & 0         & 0 \\
        d_+E_+ e^{-i \omega t}  &  \hbar\omega_0                      & 0                                             & \frac{\hbar\Omega_e}{2}         & \frac{\hbar\Omega_h}{2} \\
        d_-E_- e^{-i \omega t}  & 0                             & \hbar\omega_0                                      & \frac{\hbar\Omega_h}{2}         & \frac{\hbar\Omega_e}{2}\\
        0                               & \frac{\hbar\Omega_e}{2}                             & \frac{\hbar\Omega_h}{2} &  \hbar\omega_0  & 0 \\        
        0                               & \frac{\hbar\Omega_h}{2}                             & \frac{\hbar\Omega_e}{2} & 0         &  \hbar\omega_0
    \end{array}
    \right).
\end{equation}

Let us proceed to the linear polarizations basis with H- and V-polarized bright (1) and dark (2) excitonic states introduced as follows:

\begin{equation}
    \begin{array}{cc}
         |H1 \rangle = \frac{|+1\rangle + |-1\rangle}{\sqrt{2}},
         \,\,
         &|H2 \rangle = \frac{|+2\rangle + |-2\rangle}{\sqrt{2}},  \\
         \mathstrut
         \\
        |V1 \rangle = \frac{|+1\rangle - |-1\rangle}{\sqrt{2}},
        \,\,
        &|V2 \rangle = \frac{|+2\rangle - |-2\rangle}{\sqrt{2}}. 
    \end{array}
\end{equation}

In this linear polarizations basis $(|0\rangle, |H1\rangle, |V1\rangle, |H2\rangle, |V2\rangle)$ the Hamiltonian transforms into the following form:

\begin{equation}
    \hat{H}_{lin} =
    \left(
    \begin{array}{ccccc}
        0               & \phantom{i} (d_xE_H)^*e^{i \omega t} & -i (d_yE_V)^* e^{i \omega t}   & 0             & 0 \\
        \phantom{i}d_xE_H e^{-i \omega t}  & \hbar\omega_0   & 0               & \hbar\Omega_H               & 0 \\
        i d_yE_V e^{-i \omega t}            & 0                  & \hbar\omega_0        & 0       & \hbar\Omega_V\\
        0             & \hbar\Omega_H             & 0           & \hbar\omega_0      & 0 \\        
        0        & 0           & \hbar \Omega_V            & 0             & \hbar\omega_0
    \end{array}
    \right),
\end{equation}

where $\Omega_{H,V} = \frac{\Omega_e \pm \Omega_h}{2}$, $E_{H} = \frac{E_+ + E_-}{\sqrt{2}}$, $E_{V} = \frac{i(E_+ - E_-)}{\sqrt{2}}$. 

Energy diagram for the exciton in linear polarizations basis is shown in Fig.~\ref{FigBases}(b). The Hamiltonian could be reduced down to the $3\times3$ Hamiltonian $\hat{H}_{H}$ in the basis $(|0\rangle, |H1\rangle, |H2\rangle)$ if the system is excited only by the H-polarized light ($E_V = 0$):

\begin{equation}
    \hat{H}_{H} =
    \left(
    \begin{array}{ccc}
        0                                           & \phantom{i} d_x^*E_H^* e^{i \omega t} & 0             \\
        \phantom{i} d_xE_H e^{-i \omega t}  & \hbar\omega_0                                  & \hbar\Omega_H      \\
        0                                           & \hbar\Omega_H                                  & \hbar\omega_0      \\        
    \end{array}
    \right).
\end{equation}

It could be shown by the direct substitution into the Lindblad equation that the dephasing rates remain the same for bright and dark states in both bases.

Consequently the PE experiment with exciton addressed only by the H-polarized light (left side of the scheme in Fig.~\ref{FigBases}(b)) could serve as an example of the evolution of a three-level system examined in the text. The difference of dephasing rates of bright and dark excitons is provided at least by their different radiative decay rates, but also could have additional sources.